\documentclass[12pt]{amsart}
\usepackage{amsmath}
\usepackage{amsxtra}
\usepackage{amscd}
\usepackage{amsthm}
\usepackage{amsfonts}
\usepackage{amssymb}
\usepackage{eucal}
\usepackage{epsfig}
\usepackage{graphics}
\usepackage[dvips]{color}
\usepackage{epsf,graphics,mathrsfs,yfonts,eufrak,simplewick}
\usepackage{psfrag}

\def\({\left(}
\def\){\right)}

\newcommand{\taub}{\mbox{\boldmath$\tau$}}

\newcommand{\ds}[1]{\displaystyle #1}
\newcommand{\cb}{\mathbf{c}}
\newcommand{\bb}{\mathbf{b}}
\newcommand{\ab}{\mathbf{a}}
\newcommand{\tb}{\mathbf{t}}
\newcommand{\fb}{\mathbf{f}}
\newcommand{\kb}{\mathbf{k}}
\newcommand{\lb}{\mbox{\boldmath$\ell$}}

\newcommand{\gb}{\mathbf{g}}
\newcommand{\jb}{\mathbf{j}}

\newcommand{\nn}{\nonumber}

\newcommand{\R}{{\mathbb R}}
\newcommand{\T}{{\mathbb T}}

\newcommand{\C}{{\mathbb C}}
\newcommand{\Z}{{\mathbb Z}}

\newcommand{\cP}{\mathcal{P}}

\newcommand{\slt}{\mathfrak{sl}_2}
\newcommand{\slth}{\widehat{\mathfrak{sl}}_2}
\newcommand{\res}{{\rm res}}
\newcommand{\id}{{\rm id}}

\newcommand{\tr}{{\rm tr}}

\newcommand{\Tr}{{\rm Tr}}

\newcommand{\End}{\mathop{\rm End}}

\newenvironment{tenumerate}{
  \begin{enumerate}
  
  }{\end{enumerate}}
\newcommand{\bi}{\begin{tenumerate}}
\newcommand{\ei}{\end{tenumerate}}
\newcommand{\isoto}[1][]%
{{\mathop{\buildrel{\sim}\over\longrightarrow}\limits_{#1}}}

\def\[{\left[}
\def\]{\right]}

\newcommand{\al}{\alpha}

\newcommand{\s}{\sigma}
\newcommand{\z}{\zeta}

\numberwithin{equation}{section}

\def\rb{\mathbf{r}}
\def\xb{\mathbf{x}}

\def\J{\mathbb{J}}

\def\half{\textstyle{\frac  1 2}}

\newcommand{\bbA}{\mathbb{A}}

\def\bi{\mathbf{i}}

\begin{document}

\begin{title}[Creation operators for the Fateev-Zamolodchikov spin  chain]
{Creation operators for the Fateev-Zamolodchikov spin  chain}
\end{title}
\author{M.~Jimbo, T.~Miwa and  F.~Smirnov}
\address{MJ: Department of Mathematics, 
Rikkyo University, Toshima-ku, Tokyo 171-8501, Japan}
\email{jimbomm@rikkyo.ac.jp}
\address{TM: Department of 
Mathematics, Graduate School of Science,
Kyoto University, Kyoto 606-8502, 
Japan}\email{tmiwa@math.kyoto-u.ac.jp}
\address{FS:
Sorbonne Universit\'e, UPMC Univ Paris 06, CNRS, UMR 7589, LPTHE, F-75005, Paris, France}
\email{smirnov@lpthe.jussieu.fr}
\dedicatory{Dedicated to Ludwig Faddeev on the occasion of 
his eightieth birthday}
\begin{abstract}
In our previous works on the XXZ chain of spin one half,   
we have studied the problem of constructing 
a basis of local operators whose members have simple vacuum expectation values. 
For this purpose a pair of fermionic creation operators have been introduced. 
In this article we extend this construction to the  spin one case.  
We formulate the fusion procedure for the creation operators, and   
find a triplet of bosonic as well as two pairs of fermionic creation operators. 
We show that the resulting basis of local operators satisfies the dual reduced qKZ equation. 
\end{abstract}

\maketitle

\section{Introduction}\label{sec:Intro}

In the series of papers \cite{BK1,BK2} 
Boos and Korepin put forward a conjecture that 
the correlation functions of an infinite 
anti-ferromagnetic
XXX chain of spin 1/2
can be expressed in terms of sums with rational coefficients
of products of Riemann zeta function evaluated
at odd integer points. This provides a considerable
progress with respect to the 
multiple integral representations of \cite{JM,KMT}. 
Later the conjecture was generalised to the XXZ chain.  
Considering the inhomogeneous case also proved to be very useful \cite{BKS}. 

Proving the Boos-Korepin conjecture was the original motivation for the
series of works of H. Boos, Y. Takeyama and 
the authors \cite{BJMST1,BJMST2,HGSI,HGSII}. 
In the final paper of this series \cite{HGSIII} 
the conjecture was proved in a considerably generalised form.
Namely, we have shown that the correlation function of 
the XXZ spin chain with 
the generalised Gibbs ensemble in the sense of \cite{RDYO}
are expressed in terms of the Taylor coefficients
of two functions, depending on one and two variables. 
These functions are defined by the thermodynamic 
characteristics of the generalised Gibbs ensemble alone. 
However, we emphasise that the construction 
involved in the process of the proof is, 
in our opinion, much more important than the conjecture itself. 
Let us explain this point.

First let us fix some terminology. 
Instead of dealing with the XXZ spin chain,  
it is more convenient for us to consider 
the six-vertex model equivalent to it. 
In the latter formulation, 
the object studied in \cite{HGSIII} is the partition function on 
a cylinder carrying a certain defect,  
which we interpret as an insertion of a quasi-local operator. 
Assigning to each such insertion the corresponding partition function,  
we obtain a functional $Z$ defined on the space of quasi-local operators. 
Let us call the directrix of the cylinder the space direction, 
and the generatrix the 
Matsubara
direction. 
In the space direction we considered only spin 1/2,  
while in the Matsubara direction  
we allowed arbitrary spins and inhomogeneities.  
So, strictly speaking, 
it is not quite correct to call $Z$ the partition function 
of the six-vertex model, but we hope this should
not be a problem for the reader. 

The main problem solved in \cite{HGSII} 
consists in providing a proper description of the
space of quasi-local operators. 
It was shown that this space can be created by 
three kinds of one-parametric families of operators: 
a bosonic one, $t^*(\z)$, 
and two fermionic ones, $b^*(\z)$ and $c^*(\z)$. 
Since the bosonic operator in question is rather trivial, 
we call this the fermionic construction. 
We are mostly interested
in quasi-local operators of spin $0$.  
They are obtained by acting with an equal number
of $b^*$'s and $c^*$'s and an arbitrary number of $t^*$'s 
on the ``primary field'' $q^{2\al S(0)}$ 
(this notation will be explained in Section \ref{sec:gen}). 
For example,  
the expression
\begin{align} 
b^*(\z_1)c^*(\z_2)\bigl(q^{2\al S(0)}\bigr)\,
\label{b*c*}
\end{align}
should be considered not as one quasi-local operator, 
but as a family of quasi-local operators 
of different lengths obtained by the Taylor expansion
of \eqref{b*c*} in $\z_1^2-1$, $\z_2^2-1$. 
On the other hand the value of our functional 
$Z$ on \eqref{b*c*} is given by the
function of two variables mentioned above, 
$\omega (\z_1,\z_2)$. The situation is reminiscent of CFT: 
operators of different lengths have a 
universal formula for the expectation values. 
This led us in the years 2008--2009 
to the effort of relating the fermionic construction to CFT.

In the paper \cite{HGSIV} we have shown, together with H. Boos,  
that in the case of the homogeneous spin 1/2
Matsubara chain the fermionic construction 
allows the scaling limit, providing the fermionic
description of $c<1$ CFT. 
The functional $Z$ gives the three-point functions.
This has far-reaching consequences. 
Our construction can be generalised to the case
of an inhomogeneous spin chain in the space direction. 
In the scaling limit, this gives the one point functions 
of the sine-Gordon model. 
Then we use the results obtained for CFT in order to normalise the operators, i.e. to identify them with the CFT primary fields and descendants.
This step is necessary for the application 
of one point functions in the perturbed CFT,  
to study the short distance behaviour of multi-point correlators via OPE.
All this is explained in 
detail
in \cite{OP,HGSV}. 
Thus the fermionic construction allows us to solve the problem
of computing the one point functions in the sine-Gordon model. 
The importance of this problem was pointed out by 
Al. Zamolodchikov \cite{Alpert}. Further S. Negro and
one of the authors  showed \cite{NS1,NS2} that in 
the Liouville CFT the fermionic basis provides a neat 
solution to the reflection relations of \cite{FFLZZ}. 
This gives one more evidence in favour of the fermionic basis. 

There is one more interesting circumstance which is worth mentioning. 
This is a general property of integrable models. 
The function
$\omega (\z_1,\z_2)$ gives not 
only the value of the functional $Z$ on the family of
quasi-local operators 
\eqref{b*c*}, 
but also the value of a similar functional on one
operator for an inhomogeneous chain with 
only two space sites and the same Matsubara chain. 
This point 
has been discussed in
\cite{HGSII,HGSIII}, but we decided to
elucidate
it once again in Section 2 in a more consistent way. 

Now we come to the subject of the present paper. 
It is interesting to generalise 
our previous results to the integrable anisotropic spin 1 chain 
which was found by Fateev and Zamolodchikov \cite{FZ}
or, in our language, rather 
to the corresponding 19-vertex model. 
We avoid  using the term ``spin 1 XXZ model" because 
even being often used by specialists in integrable models, this 
term is misleading for general physics audience.
The first thing to do in this direction is to prove 
an analogue of the Boos-Korepin conjecture,
i.e., to show that the correlation functions on a finite number of sites
do not really require multiple integrals, 
 which become really boundless in this case \cite{Kitanine}. 
This was essentially done 
by A. Kl{\" u}mper, D. Nawrah and J. Suzuki
in the recent work \cite{KNS}. 
Actually the construction of \cite{HGSIII} already
allows us to treat the spin 1 representation in the Matsubara direction. 
Hence the only thing to be done
is to consider the fusion in the space direction. 
This requires proving that certain singularities cancel. 
In \cite{KNS} this was done case by case. 
On the physicists' level of rigour, the cancellation must hold 
generally because the object under the
consideration is well-defined. 
The paper \cite{KNS} gives a solid evidence that the
correlation functions in the infinite volume are expressed in terms of
certain elementary functions, so 
the zeta function is not needed in this case.

However 
we are interested in the generalisation of our previous construction. 
Namely we want a construction of the space of quasi-local operators in terms
of multi-parametric families similar to 
the one given by fermions. 
Our goals 
include the description of the super-symmetric CFT, 
the computation of 
one point functions in the super-symmetric 
sine Gordon model, and 
the solution of the reflection relations for the super-symmetric  
Liouville model. 
In the present paper we make a modest step in this direction:
we show that on the lattice the space of 
quasi-local operators 
is created by one simple boson $\tb^*(\z)$, 
four fermions $\bb^*(\z)$, $\bar{\bb}^*(\z)$
$\cb^*(\z)$, $\bar{\cb}^*(\z)$, and 
three bosons $\jb^+(\z)$, $\jb^0(\z)$, $\jb^-(\z)$.
More exactly, similarly to the spin 1/2 case, we have to take their Taylor
expansions in $\z^2-1$. 

Symbolically the spin 1 creation operators are defined 
in terms of the spin 1/2 creation operators as follows: 
\begin{align}
&\jb^+(\z)=b^*(\z q^{1/2})b^*(\z q^{-1/2})\,,\label{definitions}\\
&\jb^-(\z)=-c^*(\z q^{1/2})c^*(\z q^{-1/2})\,,\nn\\
&\jb^0(\z)=b^*(\z q^{1/2})c^*(\z q^{-1/2})+c^*(\z q^{1/2})b^*(\z q^{-1/2})\,,\nn\\
&\bb^*(\z)=b^*(\z q^{1/2})t^*(\z q^{-1/2})+b^*(\z q^{-1/2})\,,\nn\\
&\bar{\bb}^*(\z)=b^*(\z q^{1/2})+t^*(\z q^{1/2})b^*(\z q^{-1/2})\,,\nn\\
&\cb^*(\z)=c^*(\z q^{1/2})t^*(\z q^{-1/2})+c^*(\z q^{-1/2})\,,\nn\\
&\bar{\cb}^*(\z)=c^*(\z q^{1/2})+t^*(\z q^{1/2})c^*(\z q^{-1/2})\,.\nn
\end{align}
The precise sense of these formulas 
will be discussed in Section \ref{sec:fusion}.  
In fact we first define the corresponding operators for an 
inhomogeneous chain where 
independent evaluation parameters are attached to each site, 
then use them to define creation operators for the homogeneous chain. 
Our main tool
is the Russian doll construction,
see the formulae \eqref{ggg}, \eqref{defop} below.
There are two issues involved in the definition \eqref{definitions}. 
First, we must ensure that the combinations \eqref{definitions}
are well-defined. 
This concerns the case of $\jb^0(\z)$, where each term has a 
singularity and only the sum makes sense.  
Second, the product of spin 1 creation operators on the 
inhomgeneous chain have singularities on the diagonal. 
This is a new feature which was absent 
in the spin 1/2 case. 
It causes a problem in the definition of the creation operators 
for the homogeneous chain. 
We are forced to subtract the singular terms and replace 
the na{\"i}ve product by a normal ordered product. 
There is a certain arbitrariness in the choice of functions used to 
define the subtraction. 
We fix the ambiguity by demanding that, 
in the infinite volume limit in the Matsubara direction,  
all quasi-local operators constructed 
by the creation operators have vanishing expectation values. 
This requirement is needed for 
the consistency with CFT as explained in \cite{HGSIV}.

The expectation values are essentially given in terms 
of the function $\omega(\z,\xi)$. 
So the most natural way to satisfy the above requirement 
would be to compute the infinite Matsubara limit of this function. 
For that matter, some results from \cite{KNS} may be useful.  
We leave the investigation of this function 
for another publication and proceed differently here. 
It is known that in the infinite volume limit both in the Matsubara
and the space direction keeping a finite number
of inhomogeneities in space,
the density matrix is given by a solution to 
the reduced qKZ equation \cite{JM}. 
We show that our creation operators satisfy the dual reduced qKZ equation. 
This fact is interesting by itself, and it also 
allows us to fix the infinite Matsubara normalisation. 

The text is organised as follows.
In Section 2 we explain that, 
for quasi-local operators of a certain specific form, 
the computation of expectation values on an infinite homogeneous chain 
can be reduced to that on a finite inhomogeneous chain.
In Section 3, after reviewing the creation operators for 
spin 1/2, we introduce the fusion procedure to construct 
creation operators of spin 1. 
In Section 4 we show that the resulting quasi-local operators 
satisfy the dual reduced qKZ equation.
In Section 5 we give a brief summary. 
In Appendix A we review the definition and properties of 
the creation operators in the spin 1/2 case. 
In Appendix B we discuss the regularity property of the products of 
creation operators. 

Throughout this paper we fix 
$q=e^{\pi i \nu}$ where $0<\nu<1$.

\section{Functional on quasi-local operators 
in a quantum spin chain: ``infinite homogeneous''
from ``finite inhomogeneous''}\label{sec:gen}

We begin with a simple construction, which we 
shall explain for the vertex models 
related to the algebra $U_q(\slth)$, but which is actually model independent.
Our aim is to construct a family of quasi-local operators 
on an {\it infinite homogeneous} chain
using an inductive limit of monodromy matrices. 
Then, we reduce the computation of
certain functionals on this family to that of 
certain matrix elements of monodromy matrices
on a  {\it finite inhomogeneous} spin chain.
Overall in this paper we shall use the normalised trace $\Tr$, 
so that for a $d$-dimensional space we have
$$
\Tr(x)=\frac 1 d \mathrm{tr}(x)\,,
$$
$\mathrm{tr}(x)$ being the usual trace.

Consider a vertex model on a square lattice
on an infinite cylinder. 
The sites along the generatrix (we call this the {\it space direction}) 
will be counted by the index $j$, while the sites along the directrix 
(called the {\it Matsubara direction}) are counted by the index $\mathbf{m}$.
The total number of sites in the Matsubara direction is 
denoted by $\mathbf{n}$.
Consider an exactly solvable model on this cylinder whose 
Boltzmann weights are given by
$R$-matrices in the tensor product of two evaluation 
representations of $U_q(\slth)$.
We shall consider
a model which is homogeneous in the space direction 
consisting of the $d$-dimensional
representation with an equal evaluation parameter, say $1$. 
In the Matsubara direction the representations 
may be of different dimensions 
$d_{\mathbf{m}}$ and may carry different evaluation parameters
$\tau_\mathbf{m}$ (inhomogeneous model).
In the following, we will replace the role of 
the infinite homogeneous spin chain in the space direction
by an auxiliary finite inhomogeneos spin chain, 
while keeping the spin chain in the Matsubara direction.

We consider the partition function 
on the cylinder with an insertion of a
quasi-local operator 
in the space direction.
Let us give the precise definition.

Consider a finite-type quantum algebra $U_q(\slt)$  
and its affinisation $U_q(\slth)$.
We define  two spaces:
$$
\mathfrak{H}_{\mathbf{S}}
=\bigotimes\limits _{j=-\infty}^{\infty}\mathbb{C}^d\,,
\qquad 
\mathfrak{H}_{\mathbf{M}}
=\bigotimes\limits _{\mathbf{m=1}}^\mathbf{n}\mathbb{C}^{d_\mathbf{m}}\,.
$$
The first definition poses a certain problem, but we shall avoid it 
working rather with operators acting
on $\mathfrak{H}_{\mathbf{S}}$ 
than with this space itself. 
Among these operators there are well-defined quasi-local ones 
which we are going to describe. 
Denote by $H$ the Cartan generator 
of $U_q(\slt)$, and take a complex number $\al$. 
 
Denote by $H_j$ the Cartan generator acting on $V_j=\C^d$, 
the $j$-th copy of the tensor product in the space direction.
We shall consider the `primary operators':
$$
q^{\al H(0)}\,,\quad H(k)=\sum\limits_{j=-\infty}^k H_j
$$
and their descendants:
$$
q^{\al H(0)}\mathcal{O}\,,
$$
where the operator $\mathcal{O} $ acts non-trivially only on a 
finite number of sites.
We call them quasi-local operators.

Introduce another complex number $\kappa$ which
we shall couple with
$$
H(\infty)=\sum\limits_{j=-\infty}^\infty H_j\,.
$$
The partition function with an insertion of a 
quasi-local operator is defined as follows.
Consider the monodromy matrix in the Matsubara direction
$$
T_{j,\mathbf{M}}(\z)=\raisebox{.7cm}{$\curvearrowleft$} 
\hskip -.6cm\prod\limits_{\mathbf{m=1}}^{\mathbf{n}}
R_{j,\mathbf{m}}(\z/\tau_\mathbf{m})
=R_{j,\mathbf{n}}(\z/\tau_\mathbf{n})\cdots 
R_{j,\mathbf{1}}(\z/\tau_\mathbf{1})
\,,
$$
and the space-Matsubara monodromy matrix:
$$
T_{\mathbf{S},\mathbf{M}}=\raisebox{.7cm}{$\curvearrowright $} 
\hskip -.75cm\prod\limits_{j=-\infty}^{\infty}
T_{j,\mathbf{M}} (1)
=\cdots T_{-1,\mathbf{M}} (1)T_{0,\mathbf{M}} (1)T_{1,\mathbf{M}} (1)
\cdots\,.
$$
Let $T_{\mathbf{M}}(\z,\kappa)=
\mathrm{Tr}_j\bigl(T_{j,\mathbf{M}}(\z)q^{\kappa H_j}\bigr)$ 
denote the corresponding transfer matrix 
(recall that we use the normalised trace).
The main object of our study is
\begin{align}
Z^{\kappa}_{\mathbf{n}}\Bigl\{ q^{\al H(0)}\mathcal{O}  \Bigr\}
=\frac {\Tr _{\mathbf{M}}\Tr _{\mathbf{S}}\(
T_{\mathbf{S},\mathbf{M}}
\ q^{\kappa H(\infty)+\al H(0)}\mathcal{O}\)}
{\Tr _{\mathbf{M}}\Tr _{\mathbf{S}}\(
T_{\mathbf{S},\mathbf{M}}
\ q^{\kappa H(\infty)+\al H(0)}\)}
\,.
\label{partition}
\end{align}
This defines a linear functional on the space of quasi-local operators. 
We assume without loss of generality that $[H(\infty),\mathcal{O}]=0$,
since the numerator vanishes 
if $\mathcal O$ carries a non-zero spin, i.e., 
$[H(\infty),\mathcal{O}]=2s
\mathcal{O}$ where $s\neq0$.  
The values of this functional
depend not only on $\kappa$ and $\mathbf{n}$, but also on 
the dimensions of the spaces in the
Matsubara direction 
and the corresponding inhomogeneities $\tau_\mathbf{m}$. 
We suppress this dependence in order to simplify the notation. 

In the generic situation 
there are no infinities in 
$Z^{\kappa}_{\mathbf{n}}\Bigl\{ q^{\al H(0)}\mathcal{O}  \Bigr\}$. 
Indeed, suppose that the transfer matrix 
$T_{\mathbf{M}}(1,\kappa)$ 
has a unique eigenvector $|\kappa\rangle$
such that the corresponding eigenvalue $T(1 ,\kappa)$ 
has the maximal absolute value. 
Similarly let 
$\langle\kappa+\al |$ be an eigencovector 
of $T_{\mathbf{M}}(1, \kappa +\al)$ with
the eigenvalue $T(1, \kappa +\al)$ possessing the same property.
Impose the generality condition
\begin{align}
\langle \kappa+\al|\kappa\rangle\ne 0\,.
\label{general}
\end{align}

Suppose that 
$$
q^{\al H(0)}\mathcal{O}=q^{\al H(j-1)}X_{[j,k]}\,,$$ 
where $X_{[j,k]}$
acts as identity outside 
some interval $[j,k]$ ($k\ge j$). Then we say that 
$q^{\al H(0)}\mathcal{O}$ 
is supported 
on the interval $[j,k]$.
We have
\begin{align}
Z^{\kappa}_{\mathbf{n}}\Bigl\{ q^{\al H(0)}\mathcal{O}  \Bigr\}
=\frac{T(1,\kappa+\al)^{j-1}}{T(1,\kappa)^{k}}\cdot\frac {\langle \kappa+\al|\Tr _{[j,k]}\(
T_{[j,k],\mathbf{M}}
\ q^{\kappa H_{[j,k]}}
X_{[j,k]}\)|\kappa\rangle}
{\langle \kappa+\al|\kappa\rangle}
\,,
\label{partition}
\end{align}
where
$$
T_{[j,k],\mathbf{M}}=\raisebox{.7cm}{$\curvearrowright $} 
\hskip -.55cm\prod\limits_{p=j}^{k}
T_{p,\mathbf{M}} (1)\,, \quad H_{[j,k]}=\sum_{p=j}^k H_p\,.
$$
It is clear that by a simple redefinition of operators 
we can 
restrict our consideration to the case $j=1$, i.e. when $\mathcal{O}$ is supported
on the interval $[1,k]$. 
We ask ourselves a question: is there a way for constructing 
such $\mathcal{O}$, 
which is useful for the evaluation of the functional \eqref{partition} 
and which takes integrability into account?

Consider an additional evaluation representation $V_c$ of the same dimension as
$V_j$ used in the space direction.
Let us equip $V_c$ with the evaluation parameter $\z$.
Introduce the adjoint $R$-matrix
$$ 
\mathbb{R}_{c,j}(\z)(\bullet)
={R}_{c,j}(\z) \ \bullet\ {R}_{c,j}(\z)^{-1}\,.
$$
Define further the following, rather formal,  object
\begin{align}\mathbb{T}_{c,[1,\infty]}(\z)=\raisebox{.7cm}{$\curvearrowleft $} 
\hskip -.5cm\prod\limits_{j=1}^{\infty}
 \mathbb{R}_{c,j}(\z)
\,. \label{defTbb}
\end{align}
Consider a linear operator 
\begin{align*}
y_{c,[1,k]}:\End \bigl(V_1\otimes\cdots\otimes V_k\bigr)
\longrightarrow \End\bigl(V_1\otimes\cdots\otimes V_k\otimes V_c\bigr).
 \end{align*}
Our goal is to make sense of the coefficients $y^{(j)}$ in the expansion
\begin{align}
\Tr_c\(\mathbb{T}_{c,[1,\infty)}(\z)
y_{c,[1,k]}(X_{[1,k]})\)=\sum_{j=0}^\infty(\zeta^2-1)^jy^{(j)}(X_{[1,k]})
\label{makesense}
\end{align}
as an operator sending $X_{[1,k]}\in\End(V_1\otimes\cdots\otimes V_k)$
to
\begin{align*}
y^{(j)}(X_{[1,k]})\in\End \bigl(V_{[1,\infty)}\bigr)=
\lim_{l\rightarrow\infty}\End \bigl(V_1\otimes\cdots\otimes V_l\bigr).
\end{align*}
Here the limit in the right hand side is the inductive limit with respect to the inclusion
\begin{align*}
\End \bigl(V_1\otimes\cdots\otimes V_l\bigr)\otimes\id
\subset
\End \bigl(V_1\otimes\cdots\otimes V_{l+1}\bigr).
\end{align*}

\vskip .2cm
\noindent
{\bf Remark.}\quad  The above inductive limit $l\rightarrow\infty$ concerns
the right end of the interval $[1,l]$.
In 
\cite{HGSII}
this was called the right reduction. 
As for the left reduction, i.e., $k\rightarrow-\infty$
for the interval $[k,m]$, there is nothing more 
to add than the argument given in 
\cite{HGSII}.
\vskip .2cm

In what follows we shall consider functions
of $\z$ as functions of $w=\z^2-1$ near $w=0$. For example,
we understand $\z^{a}=(1+w)^{a/2}$, the branch is such that
this function equals $1$ at $w=0$.

It will be convenient to use
$$ 
\check{\mathbb{R}}_{i,j}(\z)=\mathbb{R}_{i,j}(\z)\mathbb{P}_{i,j}\,,
$$
where $\mathbb{P}_{i,j}$ stands for the adjoint action of the permutation.

We have
$$ 
\left. \check{\mathbb{R}}_{i,j}(\z)\right|_{\z^2=1}=I\,,
$$
hence
$$
\check{\mathbb{R}}_{i,j}(\z)=I+(\z^2-1)\mathbf{r}_{i,j}(\z)\,,
$$
where the operator $\mathbf{r}_{i,j}(\z)$ is regular at $\z^2=1$, 
and possesses the 
\vskip .2cm
\noindent
{\bf Property}:
$$
\mathbf{r}_{i,j}(\z)(X)=0
$$
{\it if the operator $X$ acts trivially on the $i$-th and the $j$-th 
components of the tensor product.}
\vskip .2cm
Using this property one shows, following 
\cite{HGSII},
that for any finite $l>k$
\begin{align}
&\Tr_c\(\mathbb{T}_{c,[1,l]}(\z) y_{c,[1,k]}(X_{[1,k]})\)
\label{Taylor}\\
&
=\sum_{j=k}^{l-1}(\z^2-1)^{j-k}
\rb_{j+1,j}(\z)\cdots\rb_{k+2,k+1}(\z)
\mathbb{T}_{k+1,[1,k]}(\z) (y_{k+1,[1,k]}(X_{[1,k]}))
\nn\\
&+O((\z^2-1)^{l-k})\,,
\nn
\end{align} 
as Taylor series in $\z^2-1$.

Let us emphasise the most important property of \eqref{Taylor}:
there are no gaps in the product
of successive matrices
$\rb_{j+1,j}(\z),\ldots,\rb_{k+2,k+1}(\z)$.
This is a consequence of the above Property.

So, the inductive limit $l\to\infty$ for these 
power series is well-defined: every Taylor coefficient stabilises for
sufficiently large $l$, and we define
\begin{align}
&\Tr_c\(\mathbb{T}_{c,[1,\infty)}(\z)y_{c,[1,k]}(X_{[1,k]})
\)
\label{sense1}\\&
=\sum_{j=k}^{\infty}(\z^2-1)^{j-k}
\rb_{j+1,j}(\z)\cdots\rb_{k+2,k+1}(\z)
\mathbb{T}_{k+1,[1,k]}(\z) y_{k+1,[1,k]}(X_{[1,k]})
\,.\nn
\end{align}
Notice that we can allow $y_{k+1,[1,k]}$ to depend analytically 
on $\z$ provided there is no singularity at
$\z^2=1$, i.e., if the Taylor expansion is possible. 

The above procedure can be iterated, and leads to the following construction. 
Consider auxiliary spaces 
$V_{c_1},\cdots ,V_{c_k}$ 
with the evaluation parameters $\z_1,\cdots ,\z_k$.
Consider further 
\begin{align*}
f_{c_1,\cdots ,c_k}\in \mathrm{End}(V_{c_1}\otimes \cdots\otimes V_{c_k}).
\end{align*}
Then we have a $k$-parametric generating function of local operators:
\begin{align}
&\Tr_{c_1,\cdots, c_k}\(\T_{c_1,[1,\infty]}(\z_1)\cdots \T_{c_k,[1,\infty]}(\z_k)(f_{c_1,\cdots ,c_k})\)
\label{generating function}
\end{align}

\vskip .2cm

\noindent
{\bf Remark.}\quad 
We can allow 
$f_{c_1,\cdots ,c_k}$ to depend analytically on $\z_j$ provided
they have  no singularities at $\z ^2_j=1$ and 
on the diagonals $\z_i^2=\z _j^2$, i.e., if the Taylor expansion
is possible.
\vskip .2cm

Following HGSIII, Lemma 3.1, 
one computes the functional $Z^{\kappa}_{\mathbf{n}}$ on the local operators
given by these generating functions:
\begin{align*}
&Z^{\kappa}_{\mathbf{n}}\Bigl\{ q^{\al H(0)}
\Tr_{c_1,\cdots, c_k}\(\T_{c_1,[1,\infty]}(\z_1)\cdots
\T_{c_k,[1,\infty]}(\z_k)
(f_{c_1,\cdots ,c_k})\)  \Bigr\} 
\\
&
=\frac {\langle \kappa+\al|\Tr_{c_1,\cdots, c_k}
\(
T_{c_1,\mathbf{M}}(\z_1,\kappa)\cdots T_{c_k,\mathbf{M}}(\z_k,\kappa)
\ f_{c_1,\cdots ,c_k}
\)|\kappa\rangle} 
{\langle \kappa+\al|\kappa\rangle\, 
\prod _{p=1}^k T(\z_p,\kappa)}
\,,
\nn
\end{align*}
where both left and right hand sides are 
understood
as Taylor series in $\z_j^2-1$.

So, the values of the functionals
$Z^{\kappa}_{\mathbf{n}}$ on the generating function 
\eqref{generating function}
are expressed
through the matrix elements of the products of matrices 
$T_{c_j,\mathbf{M}}(\z_j,\kappa)$. 
In other
words the computation on the {\it infinite homogeneous} chain 
$\otimes_{j=1}^\infty \C^d$
is reduced to that on the {\it finite inhomogeneous} chain 
$\otimes_{j=1}^k \C^d$
with the inhomogeneities $\z_1,\cdots ,\z_k$. 
This is a nice observation, but it becomes useful
only if we know a really good way to compute these matrix elements. 
Up to now this was known only
in the case 
where we have two-dimensional representations in the space direction. 
In the next section we shall repeat briefly this construction, 
and generalise it in order to
include the case of three-dimensional representations.



\section{Basis of quasi-local operators for spin 1 chain}\label{sec:spin1}

\subsection{The case of spin 1/2}\label{sec:review1/2} 

In the case of spin 1/2, 
a family of `good' basis $\{f_{c_1,\cdots,c_k}\}$ of finite inhomogeneous chains 
was found in \cite{HGSII}. 
In this section we briefly outline this construction. 
We shall deal with the space of quasi-local operators 
slightly more general than in the previous section, 
$$
\mathcal{W}^{(\al)}
=\bigoplus\limits_{s\in\Z} \mathcal{W}_{\al-s,s}\,,
$$
where $\mathcal{W}_{\al-s,s}$ consists of quasi-local operators 
$q^{(\al-s)H(0)}\mathcal{O}$ such that $\mathcal{O}$ has spin $s$. 
We shall also need the Cauchy-type kernel
\begin{align}
\psi(\z,\al) = \z^{\al}\frac{\z^2+1}{2(\z^2-1)}\,,
\label{psi}
\end{align}
and its `primitive' function $\Delta ^{-1}_\z\psi(\z,\al)$ 
with respect to the $q$-difference operator
$$
\Delta_{\z}f(\z)=f(q\z)-f(q^{-1}\z). 
$$
We fix possible arbitrariness in quasi-constants defining for $0<\mathrm{Re}(\al)<1/\nu$
\begin{align}
\Delta ^{-1}_\z\psi(\z,\al)
=-\frac 1 {8i}VP\int\limits _{-\infty}^{\infty}
\z^{ik+\al}\frac{\coth\frac{\pi k}{2}}{\sinh\pi\nu(k-i\al)}dk\,,
\label{Del-psi}
\end{align}
and then continuing analytically with respect to $\al$.

Consider the two-dimensional auxiliary spaces
$V_{c_j}(\z_j)$ with evaluation parameters $\z_j$, $j=1,2,\cdots$. 
Our construction is based on the following linear maps
introduced in \cite{HGSII}
$$
g_{c_k}^\epsilon (\z_k):\ 
\mathrm{End}\bigl(V_{c_1}(\z_1)\otimes\cdots \otimes  V_{c_{k-1}}(\z_{k-1})\bigr)\ 
\to\ \mathrm{End}\bigl(V_{c_1}(\z_1)\otimes 
\cdots \otimes V_{c_{k}}(\z_k)\bigr)\,,
$$
where $\epsilon_k\in \{\pm,0\}$. 
In Appendix \ref{sec:appA} we summarise
the definition of the operators $g_{c}^\epsilon(\z)$ 
and their properties. 
The operator $g_{c_k}^{\epsilon_k}(\z_k)$ depends also on the other parameters
$\epsilon_j$ and $\z_j$, $1\le j\le k-1$. 
For brevity we suppress this dependence from the notation.  
Sometimes we write the operators in a different order. For example
$$
g^{\epsilon_1}_{c_1}(\z_1)g^{\epsilon_2}_{c_2}(\z_2)(I)\,
\in \End\bigl(V_{c_2}(\z_2)\otimes V_{c_1}(\z_1)\bigr)
$$
means that we first act with an operator which `adds' the space $V_{c_2}(\z_2)$, 
and then `add' $V_{c_1}(\z_1)$ further. 
The following commutation relations hold, 
see \eqref{g-comm} and the discussions therein:
\begin{align}
g_{c_1}^{\epsilon_1}(\z_1)g_{c_2}^{\epsilon_2}(\z_2)
=(-)^{\epsilon_1\epsilon_2}
\R_{c_1,c_2}(\z_1/\z_2)
g_{c_2}^{\epsilon_2}(\z_2)g_{c_1}^{\epsilon_1}(\z_1)\,. 
\label{commg}
\end{align}

\vskip .2cm

\noindent
{\bf Remark.}\quad 
The operators $g^\epsilon_c(\z)$ used here are slightly different from those of
\cite{HGSII}. In Appendix \ref{sec:annihil}
we denote the latter by $g^\epsilon_{\mathrm{rat},c}(\z)$. 
The coefficients of the operators  $g^\epsilon_{\mathrm{rat},c}(\z)$ are
rational functions in the evaluation parameters (up to an overall power).
In contrast, those of $g^\epsilon_c(\z)$ involve 
the transcendental function \eqref{Del-psi}
(see e.g. Example 2 in Section \ref{sec:gec}). 
However it is $g^\epsilon_c(\z)$ which has a better behaviour
in relation to CFT \cite{HGSIV}, and in this paper we use it exclusively. 
\vskip .2cm

Following \cite{HGSII} introduce operators $x^{\epsilon*}(\z)$ acting on $\mathcal{W}^{(\al)}$ by 
\begin{align*}
x^{\epsilon*}(\z) (q^{(\al-s) H(0)}\mathcal{O})
 =q^{(\al-s-\epsilon) H(0)}\Tr_{c}\bigl(\mathbb{T}_{c,[n+1,\infty)}(\z)
g^\epsilon_c(\z)(\mathcal{O}_{[1,n]})\bigr),
\end{align*}
where $\mathcal{O}$ is supported on $[1,n]$ and has spin $s$. 
Usually we write
\begin{align*}
b^*(\z)=x^{+*}(\z)\,,\quad  c^*(\z)=x^{-*}(\z)\,,\quad  
t^*(\z)=x^{0*}(\z)\,.
\end{align*}
In \cite{HGSII}, \cite{HGSIII} we used boldface letters, 
but we shall reserve them for the spin 1 case in the present paper. 
One
concludes that the operators $x^{\epsilon*}(\z)$ are well defined on $\mathcal{W}^{(\al)}$ setting
$\mathbb{T}_{c,[1,k]}(\z) y_{c,[1,k]}=g^\epsilon_c(\z)$ in \eqref{sense1}.

We have the commutation relations
$$
x^{\epsilon_1 *}(\z_1) x^{\epsilon_2 *}(\z_2)
=(-)^{\epsilon_1\epsilon_2}x^{\epsilon_2 *}(\z_2) x^{\epsilon_1 *}(\z_1)\,.
$$

We construct operators on the inhomogeneous chains
by applying the Russian doll principle, i.e., recursively in the form
\begin{align}
g^{\epsilon_k}_{c_k}(\z_k)\cdots g^{\epsilon_1}_{c_1}(\z_1)(I)\,.\label{ggg}
\end{align}
It is known that the expression \eqref{ggg} is regular at $\z_j^2=1$ 
and on the diagonal $\z_j^2=\z_k^2$, $j\neq k$ (see \cite{HGSII}, Lemma 3.8).

The consecutive application of the operators $x^{\epsilon *}(\z)$ 
yields (set 
$n=0$
in  \eqref{Rdoll} )
\begin{align}
&x^{\epsilon_k *}(\z_k)\cdots x^{\epsilon_1 *}(\z_1)(q^{\al H(0)})
\label{defop}\\
&=\Tr _{c_1,\cdots ,c_k}\(\mathbb{T}_{c_1,[1,\infty)}(\z _1)\cdots 
\mathbb{T}_{c_k,[1,\infty)}(\z _k)
g^{\epsilon_k}_{c_k}(\z_k)\cdots g^{\epsilon_1}_{c_1}(\z_1)(I)\)
q^{(\al -\sum \epsilon _p)H(0)}\,.\nn
\end{align} 
We call this  ``dressed Russian doll" formula because it 
allows to define the action of operators on $\mathrm{End}\(\mathfrak{H}_\mathrm{S}\)$ by
covering the Russian doll in the auxiliary spaces $\mathrm{End}\(V_{c_1}\otimes\cdots\otimes V_{c_k}\)$ with the
adjoint monodromy matrices and tracing out the auxiliary spaces.

The virtue of \eqref{defop} is that their expectation values 
can be described explicitly. 
The main formula of \cite{HGSIII}, (1.12), 
can be written as follows
\begin{align}
&
\frac{\langle \kappa+\al|\ 
\Tr _{c_1,\cdots ,c_k}\(T_{c_1,\mathbf{M}}(\z _1)
\cdots T_{c_k,\mathbf{M}}(\z _k)
g^{\epsilon_k}_{c_k}(\z_k)\cdots g^{\epsilon_1}_{c_1}(\z_1)(I)\)
|\kappa\rangle}{\prod_{p=1}^k T(\z_p,\kappa)\langle \kappa+\al|\kappa\rangle}
\label{main}
\\
&=
\prod _{j:\epsilon _j=0}\rho(\z_j)\times 
\det \Bigl(\omega (\z_r,\z_s)\Bigr)
\Bigl|_{{r:\epsilon _r=+}\atop {s:\epsilon _s=-}}\,.
\nn
\end{align}
Here $\rho(\z)$ is a ratio of the left and right eigenvalues of the 
Matsubara transfer-matrix, 
$$
\rho (\z)=\frac {T(\z,\kappa+\al)}{T(\z,\kappa)}\,,
$$
and the function $\omega (\z,\xi)$ is given in Appendix A.  
Here we mention only the properties most essential for our goals.

The function $\omega (\z,\xi)$ consists of two pieces,
$$
\omega(\z,\xi)=\omega_\mathrm{reg}(\z,\xi)+\omega_\mathrm{trans}(\z,\xi)\,.
$$ 
The first piece $(\xi/\z)^{\al}\omega_\mathrm{reg}(\z,\xi)$ 
is a meromorphic function of $\z^2$ and $\xi^2$ with 
simple poles located at the zeros of ${T(\z,\kappa)}{T(\xi,\kappa)}$ and is 
regular 
elsewhere. 
The second piece is given by 
\begin{align}
&\omega_\mathrm{trans}(\z,\xi)=
\frac{1}{4}
\frac 1{T(\z,\kappa)T(\xi,\kappa)}
\(a(\xi)d(\z)\psi (q \z/\xi,\al)-a(\z)d(\xi)\psi (q^{-1} \z/\xi,\al) \)
\label{trans}\\
&+(1+\rho (\z)\rho(\xi))
\Delta ^{-1}_\z\psi (\z/\xi,\al)-\rho(\z)\Delta ^{-1}_\z\psi (q^{-1}\z/\xi,\al)
-\rho(\xi)\Delta ^{-1}_\z\psi (q \z/\xi,\al)\,.\nn
\end{align}
It has simple poles at $\z/\xi=q^{\pm1}$, with the residue
\begin{align}
&\res_{\z'=\z}\ \omega (q^{\pm 1/2}\z',q^{\mp1/2}\z)\frac {d{\z'}^2}{{\z'}^2}=\pm
\mathcal{N}(\z)\cdot
\frac{1}{2}
\Bigl(1+\frac {T^{(1)}(\z ,\kappa+\al)}
{T^{(1)}(\z ,\kappa)}\Bigr),
\label{res-omega}\\
&\mathcal{N}(\z)=\frac{3}{4}
\frac{T^{(1)}(\z,\kappa)}
{T(q^{-1/2}\z,\kappa)T(q^{1/2}\z,\kappa)}\,.
\label{Nfac}
\end{align}
The appearance of fused spin 1 transfer matrices 
$T^{(1)}(\z ,\kappa),T^{(1)}(\z ,\kappa+\al)$ 
in this formula is very important. 
In deriving \eqref{res-omega} we used
\eqref{trans}, 
\begin{align}
\Delta^{-1}\psi(\z,\al) =\pm \frac{1}{2}\psi(q^{\mp1}\z,\al)+O(1)
\quad (\z\to q^{\pm1}),
\label{Dinv-sing}
\end{align}
and the fusion relations for the transfer matrix eigenvalues
$$
3T^{(1)}(\z ,\kappa)
=4
T(\z q^{1/2},\kappa)T(\z q^{-1/2},\kappa)-a(\z q^{1/2})d(\z q^{-1/2})\,.
$$
(Recall again that the transfer matrices are defined via
the normalised trace).

\subsection{Fusion of spin 1/2 operators}\label{sec:fusion} 
Certainly, the road from spin 1/2 to spin 1 passes through fusion. 
In this subsection we consider the inhomogeneous chain of spin 1. 

Let us start from some generalities. 
The spin 1 representation is realised as a $U_q\bigl(\slth\bigr)$ submodule 
\begin{align*}
V_{\{1,2\}^+}(\z)\subset V_1(\z^-)\otimes V_2(\z^+)\,,
\end{align*}
where $\z^\pm=q^{\pm 1/2}\z$. 
For definiteness we normalise the $R$ matrix in such a way that 
$$
R_{1,2}(q)=D_{1,2}\cP^+_{1,2},\quad 
\res_{\z=q^{-1}}R_{1,2}(\z)\frac{d\z^2}{\z^2}
=\cP^-_{1,2}\,,
$$
where $\cP^\pm_{1,2}=(\cP^\pm_{1,2})^2$ are the projectors 
onto the symmetric and anti-symmetric subspaces,  
and $D_{1,2}$ is a diagonal matrix which commutes with $\cP^+_{1,2}$. 

Consider an operator $f_{1,2}\in \End \(V_1(\z^-)\otimes V_2(\z^+)\)$. 
If its image 
is contained in the subspace $V_{\{1,2\}^+}(\z)$, then by restriction we obtain 
\begin{align*}
f_{\{1,2\}^+}=f_{1,2}\bigl|_{V_{\{1,2\}^+}(\z)}\quad \in  \End V_{\{1,2\}^+}(\z)\,.
\end{align*}
It means that $\End V_{\{1,2\}^+}(\z)$ is a subquotient 
\begin{align}
\End V_{\{1,2\}^+}(\z)\simeq \cP^+_{1,2}\End (V_1(\z^-)\otimes V_2(\z^+) )
~\bmod~\cP^+_{1,2}\End (V_1(\z^-)\otimes V_2(\z^+) )\cP^-_{1,2}\,.
\label{End-mod}
\end{align}
More generally, let $f_{c_1,\cdots,c_{2n}}$ be an operator
on $V_{c_1}(\z^-_1)\otimes V_{c_2}(\z^+_1)\otimes
\cdots\otimes V_{c_{2n-1}}(\z^-_n)\otimes V_{c_{2n}}(\z^+_n)$,
$\z_j^\pm=q^{\pm1/2}\z_j$, and suppose that
\begin{align*}
f_{c_1,\cdots,c_{2n}} =\cP^+_{2j-1,2j}f_{c_1,\cdots,c_{2n}} \quad
(j=1,\cdots,n). 
\end{align*}
Let 
$f_{\mathbf{c_1},\cdots,\mathbf{c_{n}}}$ 
be its restriction to 
$V_{\mathbf{c}_1}\otimes\cdots\otimes V_{\mathbf{c_{n}}}$ 
where we write $\mathbf{c}_j=\{c_{2j-1},c_{2j}\}^+$ for brevity.  
From the fusion relation
\begin{align*}
R_{1,*}(q^{-1/2}\z)R_{2,*}(q^{1/2}\z)\cP^+_{1,2}=R_{\{1,2\}^+,*}(\z)
\end{align*}
we obtain the following relation between the expectation values
\begin{align}
&\frac {\langle \kappa+\al|\Tr_{c_1,\cdots, c_{2n}}
\(
T_{c_1,\mathbf{M}}(\z_1^-,\kappa)T_{c_2,\mathbf{M}}(\z_1^+,\kappa)
\cdots 
T_{c_{2n-1},\mathbf{M}}(\z_n^-,\kappa)T_{c_{2n},\mathbf{M}}(\z_n^+,\kappa)
\ f_{c_1,\cdots ,c_{2n}}\)|\kappa\rangle} 
{\langle \kappa+\al|\kappa\rangle\,
\prod _{p=1}^{n} T(\z_p^-,\kappa)T(\z_p^+,\kappa)}
\,
\label{VEV-1-1/2}
\\
&=\prod_{j=1}^n\mathcal{N}(\z_j)
\times
\frac {\langle \kappa+\al|\Tr_{\mathbf{c}_1,\cdots, \mathbf{c}_{n}}
\(
T_{\mathbf{c}_1,\mathbf{M}}(\z_1,\kappa)\cdots 
T_{\mathbf{c}_{n},\mathbf{M}}(\z_n,\kappa)
\ f_{\mathbf{c}_1,\cdots,\mathbf{c}_{n}}\)|\kappa\rangle} 
{\langle \kappa+\al |\kappa\rangle\,\prod _{p=1}^{n} T^{(1)}(\z_p,\kappa)}
\,,\nn
\end{align}
where $\mathcal{N}(\z)$ is given by \eqref{Nfac}.

Now we discuss the fusion of operators $g^\epsilon_c(\z)$. 
Consider some linear combination $\gb^{\delta}_{c_2,c_1}(\z)$
of $g^{\epsilon_2}_{c_2}(\z q^{1/2})g^{\epsilon_1}_{c_1}(\z q^{-1/2})$
where $\epsilon_1+\epsilon_2$ is fixed. In addition to (A.5--7),
it is convenient to use the operator $g^{\bar{0}}_c(\z)$ meaning
\begin{align}
g^{\bar{0}}_c(\z)(X)=X\otimes \mathrm{id}_c\,.\label{empty}\end{align}
for which the value of $\epsilon$ is understood  to be $0$.

We say $\gb^{\delta}_{c_2,c_1}(\z)$ is {\it admissible} if 
\begin{align}
\gb^{\delta}_{c_2,c_1}(\z)(I)=\cP^+_{c_1,c_2}\gb^{\delta}_{c_2,c_1}(\z)(I)\,.
\label{2sites}
\end{align}
For admissible operators the following holds:
\begin{align}
\gb^{\delta _k}_{c_{2n},c_{2n-1}}(\z_n)\cdots\gb^{\delta _1}_{c_2,c_1}(\z_1)(I)
=\cP^+_{c_{2k-1},c_{2n}}\cdots \cP^+_{c_1,c_2}
\gb^{\delta _n}_{c_{2n},c_{2n-1}}(\z_k)\cdots\gb^{\delta _1}_{c_2,c_1}(\z_1)(I)\,.
\label{2ksites}
\end{align}
To see this take $\ds{X\in \End \bigl(V_{[1,n]}\bigr)}$ 
as in Subsection A.1, 
and
note that the $R$ matrix symmetry \eqref{Rsym} implies
\begin{align}
&g^{\epsilon}_{c}
(\z_1|\xi_1,\cdots,\xi_j,q \xi_{j},\cdots,\xi_n)
\bigl(\cP^+_{j+1,j}X\bigr)
\label{gep3}
\\
&=\cP^+_{j+1,j}
g^{\epsilon}_{c}
(\z_1|\xi_1,\cdots,\xi_{j},q\xi_j,\cdots,\xi_n)
\bigl(\cP^+_{j+1,j}X\bigr)\,
\nn
\end{align}
provided the specialisation
$g^{\epsilon}_{c}(\z_1|\xi_1,\cdots,\xi_n)\bigl|_{\xi_{j+1}=q\xi_j}$
is well defined. 
Using the commutation relations \eqref{commg} we can bring  
any $\gb^{\delta _j}_{c_{2j},c_{2j-1}}(\z_k)$
to the right and use \eqref{2sites}, then 
the symmetriser goes to the left because of \eqref{gep3}.

We choose the index $\delta$ to take values in $(i,j)$, $i,j=1,2,3$. 
There are two admissible operators $\gb^{\delta}_{c_2,c_1}(\z)$ which
we can construct immediately:
$$\gb^{(3,1)}_{c_2,c_1}(\z)=
g^+_{c_2}(\z q^{1/2})g^+_{c_1}(\z q^{-1/2}),\quad 
\gb^{(1,3)}_{c_2,c_1}(\z)=
g^-_{c_2}(\z q^{1/2})g^-_{c_1}(\z q^{-1/2})\,.
$$
Indeed, $g_{c_2}^\pm(\z q^{1/2})g_{c_1}^\pm(\z q^{-1/2})(I)$ are proportional to $\sigma ^\pm_{c_2}\sigma ^\pm_{c_1}$,
which obviously satisfy \eqref{2sites}. 
These two operators look as a part of a
triplet, so we want to find 
one more operator of similar type 
which does not change the total spin. 
Certainly, it should be built of $g_{c_2}^+g_{c_1}^-$
and $g_{c_2}^-g_{c_1}^+$. 
Taking care of singularities we come with the following proposal:
\begin{align}
\gb^{(2,2)}_{c_2,c_1}(\z)=
\lim_{{\xi_1\to \z q^{-1/2}}\atop{\xi_2\to \z q^{1/2}}}
\Bigl((q\xi _1/\xi_2)^\al g^+_{c_2}(\xi_2)g^-_{c_1}(\xi_1)
+(q\xi _1/\xi_2)^{-\al} g^-_{c_2}(\xi_2)g^+_{c_1}(\xi_1)
\Bigr)\,.
\label{gb22}
\end{align}
In Appendix \ref{sec:appB} we show that the limit \eqref{gb22}
is well-defined. By a direct computation using 
\eqref{gep4}, \eqref{gep5} we find 
\begin{align}
&\gb^{(2,2)}_{c_2,c_1}(\z)(I)
\label{g22I}
\\ 
&
=-
(q^{\al}-q^{-\al})\mathcal{P}^+_{1,2}\(
\frac {q+q^{-1}}{2(q-q^{-1})}(q^\al\tau^+_{1}\tau^+_{2}
+q^{-\al}\tau^-_{1}\tau^-_{2})
+\frac {q^{\al}+q^{-\al}}{q^2-q^{-2}}
(\tau^+_{1}\tau^-_{2}+\tau^-_{1}\tau^+_{2})
\)
\nn\,. 
\end{align}
Hence it is admissible. 

Now we pass to the fermions. It is easy to see that the following combinations are admissible:
\begin{align}
&\gb^{(1,2)}_{c_2,c_1}(\z)=
g^-_{c_2}(\z q^{1/2})g^0_{c_1}(\z q^{-1/2})+g^-_{c_1}(\z q^{-1/2})\,,\nn\\
&\gb^{(2,1)}_{c_2,c_1}(\z)=
g^+_{c_2}(\z q^{1/2})g^0_{c_1}(\z q^{-1/2})+g^+_{c_1}(\z q^{-1/2})\,,\nn\\
&\gb^{(2,3)}_{c_2,c_1}(\z)=
g^0_{c_2}(\z q^{1/2})g^-_{c_1}(\z q^{-1/2})+g^-_{c_2}(\z q^{1/2})\,,\nn\\
&\gb^{(3,2)}_{c_2,c_1}(\z)=
g^0_{c_2}(\z q^{1/2})g^+_{c_1}(\z q^{-1/2})+g^+_{c_2}(\z q^{1/2})\,.\nn
\end{align}
We add two more simple operators. The unit operator for the fused chain 
$$
\gb^{(1,1)}_{c_2,c_1}(\z)=\id\,,
$$
and 
$$
\gb^{(3,3)}_{c_2,c_1}(\z)
=
g^0_{c_2}(\z q^{1/2})g^0_{c_1}(\z q^{-1/2}). 
$$
After restricting to the spin 1 subspaces we obtain
\begin{align*}
\gb^{\delta}_{\mathbf{c}_k}(\z):
\End&\bigl(V_{\mathbf{c}_1}(\z_1)\otimes \cdots\otimes
V_{\mathbf{c}_{k-1}}(\z_{k-1})\bigr)\\
&\longrightarrow
\End\bigl(V_{\mathbf{c}_1}(\z_1)\otimes\cdots\otimes 
 V_{\mathbf{c}_{k-1}}(\z_{k-1})\otimes V_{\mathbf{c}_k}(\z)\bigr)\,,
\end{align*}
where $\mathbf{c}_j=\{c_{2j-1},c_{2j}\}^+$.

\subsection{Creation operators for spin 1}\label{sec:spin1creation} 

From now on, for economy of symbols 
we use indices $c_j$ to label the three-dimensional spaces. 
There should not be a confusion with the two-dimensional case 
because for the operators we use letters of different style. 

It is easy to see that $\gb ^{(i,j)}_{c}(\z )(I)$, with $i,j=1,2,3$, 
provide nine independent operators on one site. 
Using this fact it should not be hard to prove 
the completeness for the inhomogeneous case
in the spirit of \cite{FB}.

Following our previous logic we would like now to 
define operators on an infinite chain of spin 1 from $\gb ^{(i,j)}_{c}(\z )$.
A na{\"i}ve attempt would be to use the formula 
\begin{align}
&\xb ^{(i_k,j_k)}(\z _k)\cdots \xb ^{(i_1,j_1)}(\z _1)(q^{\al H(0)})
\label{ggg2}
\\
&=\Tr _{c_1,\cdots, c_k}\bigl[
\T_{c_1,[1,\infty)}(\z_1)\cdots 
\T_{c_k,[1,\infty)}(\z_k)\gb_{c_k}^{(i_k,j_k)}(\z _k)\cdots  \gb ^{(i_1,j_1)}
_{c_1}(\z _1)(I)\bigr]\cdot q^{(\al -\sum_p(j_p-i_p))H(0)}\,.\nn
\end{align}
However, this does not work directly: the right hand side does 
not 
admit a Taylor expansion 
because of singularities on the diagonals. 
Let us explain this.

Consider for example the case
\begin{align*}
\gb^{(3,1)}_{\{3,4\}^+}(\z_2) \gb^{(1,3)}_{\{1,2\}^+}(\z_1) \bigl(q^{\al H(0)}\bigr)
=\lim g^+_4(\xi_4) g^+_3(\xi_3) g^-_2(\xi_2) g^-_1(\xi_1)
(I)\,,
\end{align*}
where we set  
\begin{align}
\xi_{2j}=q^{1/2}\z_{j}', \quad \xi_{2j-1}=q^{-1/2}\z_{j},   
\quad  (j=1,2)\,  
\label{spin1lim}
\end{align}
and take the limit $\z_j'\to\z_j$. 
In order to unveil the singularity it is easier to deal with the expectation values
than with the operators themselves. 
Take a Matsubara chain with arbitrary parameters $\taub_{\mathbf{m}}$ and 
$d_\mathbf{m}$ and compute the expectation value according to \eqref{main}
\begin{align*}
\frac {\langle\kappa+\alpha |
g^+_4(\xi_4) g^+_3(\xi_3) g^-_2(\xi_2) g^-_1(\xi_1)(I)|\kappa\rangle}
 {\langle\kappa+\alpha |\kappa\rangle}
=\omega(\xi_4,\xi_1)\omega(\xi_3,\xi_2)
-\omega(\xi_4,\xi_2)\omega(\xi_3,\xi_1)\,.
\end{align*} 
In the above limit, only the first term develops singularities. 
Let us set 
$$
\bar{f}(\z,\xi)={\textstyle \frac 1 2}
\Bigl(1+\frac{T^{(1)}(\xi,\kappa+\al)}{T^{(1)}(\xi,\kappa)}\Bigr)
\psi(\z/\xi,\al)\,
$$
and write 
\begin{align*}
&
\omega(q^{1/2}\z,q^{-1/2}\xi)=\mathcal{N}(\xi)\bar{f}(\z,\xi) + 
\omega''(\z,\xi),
\\
&\omega(q^{-1/2}\z,q^{1/2}\xi)=\mathcal{N}(\z)\bar{f}(\xi,\z) + 
\omega'(\z,\xi),
\end{align*}
where $\omega''(\z,\xi),\omega'(\z,\xi)$ are 
regular as $\z\to\xi$. 
From \eqref{res-omega} we find 
\begin{align*}
&\omega(\xi_4,\xi_1)\omega(\xi_3,\xi_2) 
=\mathcal{N}(\z_1)\mathcal{N}(\z_2)\bar{f}(\z_2,\z_1)\bar{f}(\z_1,\z_2)
\\
&+\mathcal{N}(\z_2)\bar{f}(\z_2,\z_1)
\bigl(-\omega''(\z_2,\z_1)
+\omega'(\z_2,\z_1)\bigr)+O(1)\,.
\end{align*}
On the other hand, the expectation value of $\gb^{(2,2)}(\z)$ reads
\begin{align*}
\lim_{\xi_2\to q^{1/2}\z\atop \xi_1\to q^{-1/2}\z} 
\Bigl((q\xi_1/\xi_2)^\al\omega(\xi_2,\xi_1)
-(q\xi_1/\xi_2)^{-\al}\omega(\xi_1,\xi_2)
\Bigr)
=\omega''(\z,\z)-
\omega'(\z,\z)\,.
\end{align*}
Now let us pass to the limit \eqref{spin1lim}. 
The expectation values of the spin 1 chain is related to those of spin 1/2
chain through \eqref{VEV-1-1/2}.
Comparing these, we find that the expectation value of 
\begin{align*}
(\gb^{(3,1)}_{\{3,4\}^+}(\z_2)\gb^{(1,3)}_{\{1,2\}^+}(\z_1) 
+\bar{f}(\z_2,\z_1)\bar{f}(\z_1,\z_2)-\bar{f}(\z_2,\z_1) \gb^{(2,2)}(\z_1))(I)
\end{align*}
is regular at $\z_2\to\z_1$. 

Let us return to the operators $\xb^{(i,j)}$. 
Normally we use different letters to denote them,  
\begin{align}
\bigl(\xb^{(i,j)}(\z)\bigr)_{i,j=1,2,3}=
\begin{pmatrix}
\id, &\cb ^*(\z), &\jb^-(\z)\\
\bb^*(\z),&\jb^0(\z),&\bar{\cb}^*(\z)\\
\jb^+(\z),&\bar{\bb}^*(\z), &\tb^*(\z)
\end{pmatrix}\,.
\end{align}
The formula \eqref{ggg2} does not define $\xb^{(i,j)}$ as operators acting on 
$\mathcal{W}^{(\al)}$ 
because of the singularities. However, the above investigation
of singularities suggests 
the following construction.

Choose and fix a function $f(\z,\xi)$ which is regular in the vicinity of 
$\z/\xi\in\R_{>0}$ except for a simple pole with the residue
\begin{align*}
 \res_{\z=\xi}f(\z,\xi)=\half(1+\tb^*(\xi))\,.
\end{align*}
Let us emphasise that $\tb^*(\xi)$ is in the centre of the algebra
of creation operators, hence we can manipulate it as a constant. 
Since in the above
considerations the Matsubara chain was arbitrary, we conclude that 
as an operator we have (compare \cite{COMM} where a similar argument is used)
\begin{align*}
\jb^{+}(\z)\jb^{-}(\xi)=f(\z,\xi)\jb^{0}(\xi)-f(\z,\xi)f(\xi,\z)+O(1).
\end{align*}
Notice that changing the choice of $f(\z,\xi)$ does not affect this equation
because $f(\z,\xi)f(\xi,\z)$ does not have a contribution to the simple pole term.

In the general case we have the following construction. Consider the
bosonic operators, which obey the
OPE for the current algebra
$\slth$ with the central charge $1$:
\begin{align}
&\contraction{}{\jb^{+}(\z)}{}{\jb^{-}(\xi)}\jb^{+}(\z)\jb^{-}(\xi)=
f(\z,\xi)\jb^{0}(\xi)-f(\z,\xi)f(\xi,\z)\,,
\nn\\
&\contraction{}{\jb^{0}(\z)}{}{\jb^{0}(\xi)}{\jb^{0}(\z)}{}{\jb^{0}(\xi)}=-2f(\z,\xi)f(\xi,\z)\,,
\quad
\contraction{}{\jb^{0}(\z)}{}{\jb^{\pm}(\xi)}{\jb^{0}(\z)}{}{\jb^{\pm}(\xi)}=\pm f(\z,\xi)\jb^{\pm}(\xi)\,.
\nn
\end{align}
The appearance of this algebra is as mysterious as the appearance of fermions
in the spin 1/2 case. 
Consider further two pairs of fermions $(\bb^*,\cb^*)$, $(\bar{\bb}^*,\bar{\cb}^*)$ which
transform as $\slt$-doublets, the non-trivial OPE being
\begin{align}
\
&\contraction{}{\jb^{+}(\z)}{}{\cb^{*}(\xi)}{\jb^{+}(\z)}{}{\cb^{*}(\xi)}=-f(\z,\xi)\bb^{*}(\xi)\,,\quad
\contraction{}{\jb^{-}(\z)}{}{\bb^{*}(\xi)}{\jb^{-}(\z)}{}{\bb^{*}(\xi)}=-f(\z,\xi)\cb^{*}(\xi)\,,\nn\\
&\contraction{}{\jb^{0}(\z)}{}{\bb^{*}(\xi)}{\jb^{0}(\z)}{}{\bb^{*}(\xi)}=f(\z,\xi)\bb^{*}(\xi)\,,
\quad\contraction{}{\jb^{0}(\z)}{}{\cb^{*}(\xi)}{\jb^{0}(\z)}{}{\cb^{*}(\xi)}=-f(\z,\xi)\cb^{*}(\xi)\,,\nn
\nn
&
\end{align}
and similarly for $(\bar{\bb}^*,\bar{\cb}^*)$. 
Finally, require the OPE between the two pairs of fermions:
\begin{align}
&\contraction{}{\bb^*(\z)}{}{\overline{\cb}^*(\xi)}{\bb^*(\z)}{}{\overline{\cb}^*(\xi)}=-(1-\tb^*(\xi))f(\z,\xi)\,,\quad
\contraction{}{\overline{\bb}^*(\z)}{}{\cb^*(\xi)}{\overline{\bb}^*(\z)}{}{\cb^*(\xi)}=(1-\tb^*(\xi))f(\z,\xi)\,.\nn
\end{align}
The rest of the pairs do not develop singularities in the OPE.
Using these OPE define the normal ordering $:\bullet:$ in the 
standard way. Then
the normal ordered products of our operators define families of
quasi-local operators by Taylor expansion in $\z^2-1$.
For example,
\begin{align*}
:\jb^{+}(\z)\jb^{-}(\xi):=\jb^{+}(\z)\jb^{-}(\xi)-f(\z,\xi)\jb^{0}(\xi)+f(\z,\xi)f(\xi,\z).
\end{align*}
Another example is
\begin{align*}
:\jb^0(\z)\bb^*(\xi)\bar{\cb}^*(\eta):=\{\jb^0(\z)-f(\z,\xi)+f(\z,\eta)\}
:\bb^*(\xi)\bar{\cb}^*(\eta):
\end{align*}
Let us consider the Taylor coefficient in the normal ordered expressions as
operators  $\xb^{(i,j)}_k$. 
Their commutation relations are derived as usual from the OPE.
\vskip .2cm
Let us stress that we have tautologically defined
the action of $\xb^{(i,j)}_k$ on the
Fock space created by themselves. 
We have little doubt that 
they span the entire space of quasi-local operators,   
but the proof of this statement is an unfinished task 
at this writing. 

\vskip .2cm
What is the best possible choice for the function $f(\z,\xi)$?  
Our final goal is the scaling limit which allows us to make contact 
with CFT. Namely, in the spin 1 case 
we expect to find in the limit a superconformal model with 
$c=3/2-12\nu^2/(1-2\nu )$ 
(in the spin 1/2 case, this was simply a conformal model with $c=1-6\nu ^2/(1-\nu)$).  
This must be possible if one takes 
for Matsubara the homogeneous chain of 
three-dimensional spaces with parameters $\tau_\mathrm{m}$ 
all equal to $q^{1/2}$. 
Like in the spin-1/2 case \cite{HGSIV} the 
operators $\mathbf{x} ^{\delta }(\z)$ are supposed 
to produce descendants of the primary field obtained by 
the scaling limit of $q^{\al H(0)}$. 
The one point functions of descendants
must vanish when the radius of the cylinder becomes infinite, 
i.e., when we get CFT on the plane. As explained
in \cite{HGSIV}, 
for the spin-1/2 case similar property is satisfied 
even for the lattice model if we create quasi-local 
operators by $b^*(\z)$, $c^*(\z)$, $t^*(\z)-1$. 
This means that the expectation values of these quasi-local
operators vanish for the infinite homogeneous Matsubara case. 
We wish to
retain a similar property in the spin 1 case.

Namely, we want to fix the function $f(\z,\xi)$ in such a way 
that, in the limit when the homogeneous Matsubara chain becomes infinite,  
the expectation values vanish for the descendants created by
normal ordered products of $\mathbf{x}^{\delta } (\z)$.
One way of fixing this is to compute $\omega(\z,\xi)$, 
to take the limit of infinite
homogeneous Matsubara case, and to see what  happens. 
We would like to leave all such computations for 
another paper, but we shall comment later on the 
results from the related paper \cite{KNS}.
We shall take another way, which is interesting by itself,  
to fix the normalisation of operators in the infinite homogeneous chain.
It is known \cite{JM} that the density matrix for
the infinite homogeneous Matsubara case is given by 
special solutions to the reduced 
quantum Knizhnik-Zamolodchikov
(rqKZ) equations. 
We shall check the
quasi-local operators created by $\mathbf{x}^{\delta }(\z)$ 
against this description of the density matrix in the next section.

\section{Reduced qKZ equations}\label{sec:rqKZ}
Let us consider in general a model on a lattice which is infinite in both
directions,  
whose rows and columns correspond to the same 
$d$-dimensional evaluation representation $V^{(d)}$
of $U_q(\slth)$.
The rqKZ
equation 
is a general way of calculating correlation functions 
in this situation \cite{JM}. 
Assume that the evaluation parameters are 
$\tau_\mathbf{m}=q^{1/2}$ 
for all rows and $1$ for all columns,  
except for $n$ consecutive columns to which 
independent parameters $\z_1,\cdots, \z _n$ are attached. 
Insert a local operator $X_{[1,n]}$ at these lines, 
and put a semi-infinite tail $q^{\al H(0)}$ 
(see \cite{JM} for more explication and graphical representation). 
The density matrix $h(\z_1,\z_2,\cdots ,\z_n )$ is a linear functional 
which associates to $X_{[1,n]}$ the partition function 
\begin{align}
h(\z_1,\z_2,\cdots ,\z_n )\bigl(X_{[1,n]}\bigr)
=\lim
\frac{\langle \kappa+\al|\ 
\Tr _{1,\cdots ,n}\( T_{1,\mathbf{M}}(\z _1)\cdots 
T_{n,\mathbf{M}}(\z _n)
X_{[1,n]}\)|\kappa\rangle}{\prod_{p=1}^nT(\z_p,\kappa)\langle \kappa+\al|\kappa\rangle}\,,
\nn
\end{align}
where $\lim $ stays for the limit of infinite homogeneous Matsubara chain.
It satisfies the rqKZ equation
\begin{align}
& h(\z_1,\cdots ,\z_j q ^{-1},\cdots \z_n)
= h(\z_1,\z_2,\cdots ,\z_n )\circ 
\bbA_{j,[1,n]}(\z_1,\cdots,\z_n)\,,
\label{rqKZ}
\end{align}
where $\bbA_{j,[1,n]}(\z_1,\cdots,\z_n)$ is defined as follows. 
\begin{align}
&\bbA_{j,[1,n]}(\z_1,\cdots,\z_n)(X_{[1,n]})
\label{defA}
\\
&\quad=T_{j,[1,j-1]}q^{\al H_j}
\theta_j\Bigl(\theta _j(T_{j,[j+1,n]}^{-1})X_{[1,n]}\ 
\theta _j(T_{j,[1,j-1]}^{-1})\Bigr)T_{j,[j+1,n]}
\,, \nn
\\
&T_{j,[k,m]}=R_{j,m}(\z_j/\z_m)\cdots R_{j,k}(\z_j/\z_k)\,.\nn
\end{align}
Here 
$\theta(X)=C\cdot {}^tX\cdot C^{-1}$, 
with $C:V^{(d)*}(q\z)\to V^{(d)}(\z)$ being 
an isomorphism of $U_q(\slth)$ modules. 
For practical calculations it is convenient to rewrite \eqref{defA} 
introducing two auxiliary spaces $V^{(d)}_a(\z_j)$ and $V^{(d)}_b(\z_j)$:
\begin{align*}
&\bbA_{j,[1,n]}(\z_1,\cdots,\z_n)(X_{[1,n]})
\\
&=T_{j,[1,j-1]}q^{\al H_j}\tr_{a,b}\Bigl(P_{b,j}
T_{a,[j+1,n]}^{-1}\theta_j (X_{[1,n]})\ 
T_{b,[1,j-1]}^{-1}P_{a,j}\Bigr)T_{j,[j+1,n]}
\end{align*}
Let us mention some  general properties  of \eqref{defA}: 
\begin{align}
&\bbA_{n-1,[1,n-1]}(\z_1,\cdots,\z_{n-1})(X_{[1,n-1]})
=\bbA_{n-1,[1,n]}(\z_1,\cdots,\z_n)(X_{[1,n-1]})\,,
\label{redA}\\
&\bbA_{n,[1,n]}(\z_1,\cdots,\z_n)(X_{[1,n-1]})
=\T_{n,[1,n-1]}(\z_n)(q^{\al H_n}X_{[1,n-1]})\,,
\label{1dif1}\\
&\bbA_{n,[1,n]}(\z_1,\cdots,\z_n)
(\T_{n,[1,n-1]}(\z_n q^{-1})(q^{\al H_n}X_{[1,n-1]}))
=X_{[1,n-1]}\,.\label{1dift}
\end{align}

In addition to \eqref{rqKZ}, 
the density matrix has the following characteristic properties.

\vskip .2cm
\noindent 1. In the limit $\z_n^{\pm 1}\to\infty$,  $h(\z_1,\z_2,\cdots ,\z_n )$ reduces to
$h(\z_1,\z_2,\cdots ,\z_{n-1} )$:
\begin{align} 
&h(\z_1,\z_2,\cdots ,\z_n )(X_{[1,n-1]}\otimes x_n)\label{inf1}
\\
&\quad\to\ h(\z_1,\z_2,\cdots ,\z_{n-1} )(X_{[1,n-1]})
\frac{\Tr _n\bigl(q^{-\al H_n/2 }x_n\, \bigr)}
{\Tr _n\bigl(q^{-\al H_n/2 }\bigr)}\,.\nn
\end{align}

\vskip .2cm
\noindent
2. $h(\z_1,\z_2,\cdots ,\z_n )$ is regular on the diagonals $\z_i=\z_j$, 
$i\neq j$.
\vskip .2cm

Let us return to the model with spin 1/2. 
In this case we have a distinguished 
basis of $\End \(V(\xi_1)\otimes\cdots\otimes V(\xi_n)\)$,  
\begin{align*}
g^{\epsilon _n,\cdots,\epsilon_1}(\xi _n,\cdots, \xi _1)
=g_n^{\epsilon _n}(\xi_n)\cdots g_1^{\epsilon_1}(\xi _1)(I)\,
\quad (\epsilon_1,\cdots,\epsilon_n\in\{0,\bar{0},+,-\})
\end{align*}
which satisfies the dual rqKZ equation \eqref{dualrqKZ}
(recall the definition \eqref{empty}, and
see Subsection \ref{sec:rqKZ1/2} for more explanations). 

The roles of the rqKZ and the dual rqKZ equations are quite different. 
The rqKZ equation considered here is closely related to 
the qKZ equation `of level $-4$'. 
We are interested in one particular solution which describes 
the density matrix. 
In contrast, the dual rqKZ equation 
is closely related to 
the qKZ equation `of level $0$'. 
It describes a good basis of the space of operators,  in much the same way 
as in the original form factor equations \cite{book}.


Consider the pairing
\begin{align}
U^{\epsilon_1,\cdots,\epsilon _n}(\xi _1,\cdots ,\xi _n)=
 h(\xi _1,\cdots ,\xi _n)
\Bigl(g^{\epsilon _n,\cdots,\epsilon_1}(\xi _n,\cdots, \xi _1)\Bigr)\,,
\label{couple}
\end{align}
where $\epsilon _j=\pm,0, \bar{0}$. 
The equations \eqref{rqKZ}, \eqref{dualrqKZ}
result in the following simple difference equation:
\begin{align}
U^{\epsilon_1,\cdots,\epsilon_j,\cdots,\epsilon _n}(\xi _1\cdots,\xi_jq^{-1},\cdots ,\xi _n)=(-1)^{\epsilon_j}
U^{\epsilon_1,\cdots,\overline{\epsilon_j},\cdots,\epsilon _n}(\xi _1\cdots,\xi_j,\cdots ,\xi _n)\,.
\end{align}
Suppose that $\xi_1,\cdots \xi _{j-1}, \xi _{j+1},\cdots \xi_n$ are real. Then 
$U^{\epsilon_1,\cdots,\epsilon_j,\cdots,\epsilon _n}(\xi _1\cdots,\xi_j,\cdots ,\xi _n)$ is a regular 
function of $\xi _j^{1/\nu}$ since there are no singularities for $\arg(\xi _j)\le \pi\nu/2$.
This function is odd  if $\epsilon _j=\pm$, while for  $\epsilon _j=0, \bar 0$ it may contain
even and odd parts.
Further it does not vanish identically only if $\sum\epsilon_p=0$.
Consider for definiteness the case $0<\al<2$.
Suppose $\epsilon _j=+$. Then it can be shown that 
\begin{align}
U^{\epsilon _1,\cdots,\epsilon_n}(\xi _1,\cdots, \xi _n)
=\begin{cases}
O(1) & \text{for\ \  $\xi _j\to\infty$},\\
o(1) & \text{for\ \  $\xi _j\to 0$}\,.
\end{cases} \label{inf2}
\end{align}
The example \eqref{gep4} is instructive, actually it reflects well the generic situation.
On the other hand
it is easy to see that for $\epsilon _j=0, \bar 0$
\begin{align}
U^{\epsilon _1,\cdots,\epsilon_n}(\xi _1,\cdots, \xi _n)
\to 
U^{\epsilon _1,\cdots,\widehat{\epsilon_j},\cdots,\epsilon_n}(\xi _1,\cdots,\widehat{\xi_j},\cdots, \xi _n) \,,\quad  \text{for $\xi _j\to0,\ \infty$}\,.\label{0bar0}
\end{align}
From \eqref{inf2} and \eqref{0bar0} we conclude that 
$U^{\epsilon_1,\cdots,\epsilon _n}(\xi _1,\cdots,\xi _n)$ vanishes if 
at least one of $\epsilon _j$ is  $+$, 
and equals to 1 otherwise, i.e. when all $\epsilon_j$ equal $0$ or $\bar 0$.
This means that for the 
infinite homogeneous spin 1/2 Matsubara spin chain we have
$$
\rho_\infty (\z)=1,\quad \omega _\infty(\z,\xi)=0\,.
$$
This is an important property of the creation operators 
$b ^*(\z),c^*(\z),t^*(\z)$. 

Now we want to test the fusion.
In order to avoid confusions we put a superscript $1/2$ or $1$  
to the operator $\bbA_{{j,[1,n]}}$ for spin 1/2 or of spin $1$.  
Using \eqref{1de} and \eqref{2de} we obtain
\begin{align}
&g^{\epsilon}_{ 2n}(\xi _{ 2n})g^{\epsilon'}_{{ 2n}-1}(\xi _{{ 2n}-1})(X_{[1,{ 2n}-2]})
\\
&=g^{\epsilon}_{ 2n}(\xi _{ 2n})\bbA^{(1/2)}_{{ 2n}-1,[1,{ 2n}-1]}
g^{\bar{\epsilon}'}_{{ 2n}-1}(\xi _{{ 2n}-1} q^{-1})(X_{[1,{ 2n}-2]})\nn\\
&=\bbA^{(1/2)}_{{ 2n}-1,[1,{ 2n}]}\Bigl[g^{\epsilon}_{{ 2n},[1,{ 2n}-1]}
(\xi_{ 2n})
\Bigr]
_{\xi _{{ 2n}-1}\to \xi _{{ 2n}-1} q^{-1}}
g^{\bar{\epsilon}'}_{{ 2n}-1}(\xi _{{ 2n}-1} q^{-1})(X_{[1,{ 2n}-2]})
\nn\\
&=
\bbA^{(1/2)}_{{ 2n}-1,[1,{ 2n}]}\Bigl[\bbA^{(1/2)}_{{ 2n},[1,{ 2n}]}\Bigr]
_{\xi _{{ 2n}-1}\to \xi _{{ 2n}-1} q^{-1}}\nn
\\
&\times
\Bigl[g^{\bar{\epsilon}}_{{ 2n},[1,{ 2n}-1]}(\xi_{ 2n})
g^{\bar{\epsilon}'}_{{ 2n}-1}(\xi _{{ 2n}-1} )(X_{[1,{ 2n}-2]})
\Bigr]_{\xi _{{ 2n}-1}\to \xi _{{ 2n}-1}q^{-1}, \xi _{{ 2n}}\to 
\xi _{{ 2n}} q^{-1}}\,.\nn
\end{align}
It is convenient to present the result of computation of the 
product of operators $\mathbb{A}$ introducing two auxiliary spaces, 
\begin{align}
&\bbA^{(1/2)}_{{ 2n}-1,[1,{ 2n}]}\Bigl[\bbA^{(1/2)}_{{ 2n},[1,{ 2n}]}
\Bigr]
_{\xi _{{ 2n}-1}\to \xi _{{ 2n}-1} q^{-1}}(Y_{[1,{ 2n}]})
\nn\\
&=
T_{{ 2n}-1,[1,{ 2n}-2]}T_{{ 2n},[1,{ 2n}-2]}
q^{\al (\sigma^3_{{ 2n}-1}+\sigma^3_{ 2n})}
\nn\\
&\times
\mathrm{tr}_{a,b}
\Bigl(P_{b,{ 2n}}P_{a,{ 2n}-1}
R_{{ 2n},{ 2n}-1}(\xi_{ 2n}/\xi _{{ 2n}-1})
\theta _{{ 2n}-1}\theta _{{ 2n}}(Y_{[1,{ 2n}]})
T^{-1}_{b,[1,{ 2n}-2]}T^{-1}_{a,[1,{ 2n}-2]}\Bigr)
R_{{ 2n}-1,{ 2n}}(\xi _{{ 2n}-1}/\xi_{ 2n})\,.\nn
\end{align}
Consider now the limit 
$\xi_{ 2n}=\z _nq^{1/2+\varepsilon},\ 
\xi_{{ 2n}-1}=\z_n q^{-1/2}$ with $\varepsilon\to 0$. 
Obviously  
$R_{{ 2n}-1,{ 2n}}(\xi _{{ 2n}-1}/\xi_{ 2n})$ poses a problem 
because it contains a singularity. 
Suppose, however,
that in this limit
\begin{align}Y_{[1,{ 2n}]}
= \cP^+_{{ 2n}-1,{ 2n}}Y_{[1,{ 2n}]}
+Z_{[1,{ 2n}]}\cdot \varepsilon+o(\varepsilon)\,,
\label{cond}\end{align}
with some $Z_{[1,{ 2n}]}$.
Then 
\begin{align}
&\theta _{{ 2n}-1}\theta _{{ 2n}}(Y_{[1,{ 2n}]})R_{{ 2n}-1,{ 2n}}(\xi _{{ 2n}-1}/\xi_{ 2n})
\nn\\
&= \theta _{{ 2n}-1}\theta _{{ 2n}}(Y_{[1,{ 2n}]})\cP^+_{{ 2n}-1,{ 2n}}+
\theta _{{ 2n}-1}\theta _{{ 2n}}(Z_{[1,{ 2n}]})\cP^-_{{ 2n}-1,{ 2n}}
+o(1)\,.\nn
\end{align}
Now the limit exists, 
we replace $R_{{ 2n},{ 2n}-1}(\xi_{ 2n}/\xi _{{ 2n}-1})$ 
by $R_{{ 2n},{ 2n}-1}(q)$, the latter contains
$\cP^+_{{ 2n}-1,{ 2n}}$ which starts to travel giving finally
\begin{align}
&\bbA^{(1/2)}_{{ 2n}-1,[1,{ 2n}]}\Bigl[\bbA^{(1/2)}_{{ 2n},[1,{ 2n}]}\Bigr]
_{\xi _{{ 2n}-1}\to \xi _{{ 2n}-1} q^{-1}}(Y_{[1,{ 2n}]})\label{xx}
\\
&\to 
\cP^+_{{ 2n}-1,{ 2n}}\Bigl(\mathbb{A}^{(1)}_{\{{ 2n}-1,{ 2n}\}^+,[1,{ 2n}-2]}
(Y_{[1,{ 2n}]})\cP^+_{{ 2n}-1,{ 2n}}+\mathbb{A}^{(1)}_{\{{ 2n}-1,{ 2n}\}^+,[1,{ 2n}-2]}
(Z_{[1,{ 2n}]})\cP^-_{{ 2n}-1,{ 2n}}\Bigr)\,.
\nn
\end{align}
Now recall that our combinations $\gb^\delta_{2n,2n-1}(\z)$ 
are exactly 
such that $$Y_{[1,{ 2n}]}=\gb^\delta_{{ 2n},{ 2n}-1}(\z _nq^{-1})
(X_{[1,{ 2n}-2]})$$ satisfies the condition \eqref{cond}.
Moreover, 
in view of \eqref{End-mod} the last term in \eqref{xx} does not count. 
So, we conclude that
\begin{align}
\gb^\delta_{n,n-1}(\z )(X_{[1,n-1]})
=(-1)^\delta \bbA^{(1)}_{n,[1,n]} 
\gb^{\bar{\delta}}_{n,n-1}(\z q^{-1})(X_{[1,n-1]})\,,
\label{qKZspin1}
\end{align}
here and later we set
$$(-1)^{(i,j)}=(-1)^{i+j}\,,\qquad 
\overline{(i,j)}=(4-j,4-i)\,.
$$

Now we proceed in the same way as for the spin-1/2 case, namely, we couple the solution to
rqKZ equation with the operator created by 
$\gb^{{\delta}_j}(\z_j)$ (the index $j$ 
refers to three-dimensional space now):
\begin{align}
U^{\delta_1,\cdots,\delta _n}(\z _1,\cdots ,\z _n)=
 h(\z _1,\cdots ,\z _n)
\Bigl(\gb_n^{\delta_n}(\z_n)\cdots \gb_1^{\delta_1}(\z_1)(I)\Bigr)\,.
\label{couple1}
\end{align}
As before this function satisfies the equation
\begin{align}
U^{\delta_1,\cdots\delta_j ,\cdots,\delta _n}(\z _1,\cdots ,\z _j q^{-1},\cdots \z _n)=(-1)^{\delta_j}
U^{\delta_1,\cdots\overline{\delta_j },\cdots\delta _n}(\z _1,\cdots ,\z _j,\cdots \z _n)\,.\label{jjj}
\end{align}
However, contrary to the spin-1/2 case we cannot conclude immediately that 
\eqref{couple1} 
is either $0$ or $1$, since
the operator created by $\gb^{\delta}_j(\z_j)$ contains singularities on the diagonals. 
Following the discussion at the end of the previous section we want to eliminate them
by normal ordering. But we do not want to spoil the difference relation. This will be achieved
if we set
\begin{align}
f(\z,\xi)=(1+\tb^*(\xi))\frac {2(\z\xi)^{\frac 1 {\nu}}}{\nu(\z ^{\frac 2 {\nu}}-\xi ^{\frac 2 {\nu}})}\,.\label{deff}
\end{align}
Notice that we put a function 
anti-periodic under $\z\to \z q$ 
in order to preserve the signs for fermions and bosons in
\eqref{jjj}. 
Plugging in \eqref{couple1} 
the normal ordered product of $\gb^{{\delta}_j}(\z_j)$ we obtain
$0$ if some ${\delta}_j$ differ from $(1,1)$, $(3,3)$, and $1$ otherwise. 

\section{Summary and discussions}\label{sec:summary}

We have shown how to construct the operators $\bb^*(\z)$, $\cb^*(\z)$, $\bar{\bb}^*(\z)$, $\bar{\cb}^*(\z)$, $\jb^\pm(\z)$, $\jb^0(\z)$ and 
$\tb^*(\z)$ which create the quasi-local operators for the
Fateev-Zamolodchikov 
spin chain. The linear functional $Z^{\kappa}_\mathbf{n}$ can be
evaluated on these quasi-local fields using the formulae \eqref{definitions} from Introduction,
but certain adjustment is needed. Namely, 
from the main text it is clear that we have to 
 multiply formally the right hand side
of \eqref{definitions} by $\mathcal{N}(\z)^{-1}$ in order to
change the normalisation of $Z^{\kappa}_\mathbf{n} $ to the one appropriate for the spin 1 case.
Let us clarify this giving one example of the final formulae:
$$
Z^{\kappa}_\mathbf{n}\{\jb^+(\z_1)\jb^-(\z_2)(q^{\al H(0)})\}
=\frac 1 {\mathcal{N}(\z_1)\mathcal{N}(\z_2)}\left|
\begin{matrix}\omega(\z_1q^{1/2},\z_2  q^{1/2})
&\omega(\z_1q^{1/2},\z_2  q^{-1/2})\\
\omega(\z_1q^{-1/2},\z_2  q^{1/2})
&\omega(\z_1q^{-1/2},\z_2  q^{-1/2})\end{matrix}\right|\,.
$$
As has been explained  $\jb^+(\z_1)\jb^-(\z_2)(q^{\al H(0)})$ does not define a two-parametric
family of quasi-local operators. It has to be normal ordered. The quasi-local operators
are created by
$$
\jb^+(\z_1)\jb^-(\z_2)-\jb^0(\z_2)f(\z_1,\z_2)-f(\z_1,\z_2)f(\z_2,\z_1)\,,
$$
and the value of $Z^{\kappa}_\mathbf{n}$ on 
$\jb^0(\z)(q^{\al H(0)})$ is given by
$$ Z^{\kappa}_\mathbf{n}\{\jb^0(\z)(q^{\al H(0)})\}
=\frac 1 {\mathcal{N}(\z)}
\bigl[ \omega(\z q^{1/2},\z  q^{-1/2}) -\omega(\z q^{-1/2},\z  q^{1/2})\bigr] _{\mathrm{reg}}\,,
$$
following the regularisation  in  the definition \eqref{gb22}.

In principle this kind of formulae allows us 
to evaluate $Z^{\kappa}_\mathbf{n}$ on any
quasi-local operator. However, we do not think that 
for the operators of small length this way of doing adds much to the direct fusion computation of 
\cite{KNS}. We rather consider the result of this paper as an existence theorem of a good
basis in the space of quasi-local operators. We hope to be able to use this basis for
considering the scaling limit which includes the super-symmetric
CFT and its massive deformation, super sine-Gordon model. From that point of view  the present
paper should be considered as a preparation for this future investigation.

\appendix

\section{Review of the spin 1/2 case}\label{sec:appA}

In this appendix we give a summary of the definition and the properties of the 
operators $g^\epsilon_c(\z)$ used in the main text.  In particular,
we give a proof of the dressed Russian doll formula.

\subsection{Operators $g^\epsilon_c(\z)$}\label{sec:gec}

Fix an interval $[1,n]\subset\Z$.  
For $j\in[1,n]$, denote by $V_j(\xi_j)$ the two-dimensional representation of 
$U_q(\slth)$ with evaluation parameter $\xi_j$, 
and set 
\begin{align*}
V_{[1,n]}=V_1(\xi_1)\otimes\cdots\otimes V_{n}(\xi_n)\,.
\end{align*}
We shall deal with several linear operators on $\ds{\End \bigl(V_{[1,n]}\bigr)}$. 

Let us start from an adjoint action by monodromy matrices. 
Let $W_{*}(\z)$ be an `auxiliary space' (i.e., 
a representation of the Borel subalgebra of $U_q(\slth)$), 
and let $L_{\ast,j}(\z/\xi_j)$ be
the image of the universal $R$ matrix on 
$\ds{W_{\ast}(\z)\otimes V_j(\xi)}$. 
For $\ds{X\in \End \bigl(V_{[1,n]}\bigr)}$, we set 
\begin{align*}
&\T_{\ast,[1,n]}(\z)(X)=
T_{\ast,[1,n]}(\z)\cdot X\cdot T_{\ast,[1,n]}(\z)^{-1}\,,
\\
&T_{\ast,[1,n]}(\z)=L_{\ast,n}(\z/\xi_n)\cdots L_{\ast,1}(\z/\xi_1)\,.
\end{align*}
As for the auxiliary space $W_{*}(\z)$, various choices are possible.   
The ones relevant to us are the two dimensional module 
$W_{*}(\z)=V_{a}(\z)$, and
the tensor product module
$W_{*}(\z)=\ds{V_a(\z)\otimes V_A(\z)}$ 
where $V_A(\z)$ stands for the representation 
of the Borel subalgebra by $q$-oscillators
\begin{align*}
q^{-1}\ab\ab^*-q\, \ab^*\ab=q^{-1}-q,\quad
q^D\ab^* q^{-D}=q\,\ab^*,\quad  
q^D\ab\, q^{-D}=q^{-1}\ab\,.
\end{align*}
For the formulas of the corresponding monodromy matrices
$\ds{\T_{a,[1,n]}(\z)}$ and 
$\ds{\T_{\{a,A\},[1,n]}(\z)}$, 
the reader is referred to \cite{HGSII}, eq.(2.4) and eq.(2.19).  

Next we introduce operators $\kb(\z),\lb(\z),\fb(\z)$ on 
$\End \bigl(V_{[1,n]}\bigr)$ 
by certain traces involving monodromy matrices. 
Their definition depends on the spin of the operand
$\ds{X\in \End \bigl(V_{[1,n]}\bigr)}$.   
Here we say that $X$ has spin $s\in\Z$, and write as 
$s\bigl(X\bigr)=s$,
if $\ds{[\half H_{[1,n]},X]=s X}$ holds with 
$\ds{H_{[1,n]}=\sum_{j=1}^n \s^3_j}$.  

For $X\in\End \bigl(V_{[1,n]}\bigr)$ 
with $s\bigl(X\bigr)=s$, define
\begin{align}
&\kb(\z)(X)=\z^{\al-2s-1}\Tr_{a,A}\Bigl\{\s^+_a\T_{\{a,A\},[1,n]}(\z)
q^{(\al-s-1)(2D_A+\s^3_a)}\bigl(q^{-H_{[1,n]}}X\bigr)
\Bigr\}\,,
\label{kbar}\\
&\lb(\z)(X)=\z^{\al-2s-1}\Tr_{a,A}\Bigl\{\bigl(\s^3_a-\ab_A\s^+_a
\bigr)
\T_{\{a,A\},[1,n]}(\z)
q^{(\al-s-1)(2D_A+\s^3_a)}\bigl(q^{-H_{[1,n]}}X\bigr)
\Bigr\}\,,
\label{lbar}\\
&\fb(\z)(X)=\Delta_\z^{-1}\kb(\z)(X)\,.
\label{fbar}
\end{align}
In the last line the symbol $\Delta_\z^{-1}$ is defined by the integral
\begin{align*}
\Delta_\z^{-1}\kb(\z)(X)
=\int_{\Gamma}\Delta_\z^{-1}\psi(\z/\xi,\al)\cdot
\kb(\xi)(X)\frac{d\xi^2}{2\pi i \xi^2}\,,
\end{align*}
where the contour $\Gamma$ is such that 
the poles of $\Delta^{-1}_{\z}\psi(\z/\xi,\al)$
are outside and the poles of $\kb(\xi)(X)$ are inside.   
The function $\xi^{-\al}\kb(\xi)(X)$ is rational in $\xi^2$,  
behaves as $O(\xi^{-2})$ at $\xi^2=\infty$,  
and has poles only at $q^2\xi^2_j,\xi^2_j,q^{-2}\xi^2_j$ ($j\in [1,n]$) and at $0$
when $s>0$.

Finally, for $\epsilon=0, \bar0 ,+,-$ we define the linear maps 
\begin{align}
 g^\epsilon_{c}(\z)~:~
\End \bigl(V_{[1,n]} \bigr)
\longrightarrow \End \bigl( V_{[1,n]}\otimes V_c(\z)\bigr)\,\label{gep1}
\end{align}
as follows. Assuming $s\bigl(X\bigr)=s$, we set
\begin{align}
g^0_{c}(\z)(X)
&=\T_{c,[1,n]}(\z)\left(X
\cdot q^{(\al-s) \s^3_c}\right)\label{g0} \,,\\
g^{\bar0}_{c}(\z)(X)&=X\otimes{\id}\label{g0bar}\,,\\
g^+_{c}(\z)(X)
&=\half \fb(q\z)(X)+\half \fb(q^{-1}\z)(X)
-g^0_{c}(\z)\Bigl(\fb(\z)(X)\Bigr)\label{g+}\\
& -\half\s^3_c\cdot \kb(\z)(X)+\s^+_c\cdot \lb(\z)(X)\nn\,,
\\
g^-_{c}(\z)(X)&=-q^{-\al+2s-2}(1-q^{2(\al-2s+1)})
\times \Bigl\{\Bigl(\J\circ g^+_{c}(\z)\circ\J\Bigr)
(X)\Bigr\}\Bigl|_{\al\to-\al+2s}\,.\label{g-}
\end{align}

When necessary, we exhibit explicitly 
the dependence on the evaluation parameters and write \eqref{gep1} as 
\begin{align*}
g^{\epsilon}_{c,[1,n]}(\z|\xi_1,\cdots,\xi_n)\,.
\end{align*}
In this notation, for example, the composition 
\begin{align*}
\End \bigl(V_{[1,n]} \bigr)\longrightarrow 
\End\bigl(V_{[1,n]}\otimes V_{c_1}(\z_1)\bigr)  
\longrightarrow 
\End \bigl(V_{[1,n]}\otimes V_{c_1}(\z_1) \otimes V_{c_2}(\z_2)\bigr)\,
\end{align*}
is given by 
\begin{align}
g^{\epsilon_2}_{c_2,[1,n]\cup c_1}(\z_2|\xi_1,\cdots,\xi_n,\z_1)\circ
g^{\epsilon_1}_{c_1,[1,n]}(\z_1|\xi_1,\cdots,\xi_n)\,.
\label{g-comp}
\end{align}
To unburden the notation 
we often abbreviate \eqref{g-comp} to 
$\ds{g^{\epsilon_2}_{c_2}(\z_2)g^{\epsilon_1}_{c_1}(\z_1)}$, 
keeping in mind that the space $V_{c_1}(\z_1)$ is `added' first 
and  $V_{c_2}(\z_2)$ the next. 

\bigskip

\noindent{\it Example 1.}\quad 
In the simplest case when $[1,n]=\emptyset$, 
the action of $g^\epsilon_c(\z)$ on the unit operator $I$ reads
\begin{align*}
&g^0_{c}(\z)(I)
=q^{\al \s^3_c},\\
&g^+_{c}(\z)(I)
=-(q^{-1}\z)^{\al-1}\s^+_c, 
\quad
g^-_{c}(\z)(I)
=q^{-1}(1-q^{2(\al+1)})\z^{-\al-1}\s^-_c\,.
\end{align*}
Here and after, 
\begin{align*}
\tau^+=\begin{pmatrix}
	1 & 0 \\ 0 & 0 \\
       \end{pmatrix}\,,
\quad
\tau^-=\begin{pmatrix}
	0 & 0 \\ 0 & 1 \\
       \end{pmatrix}\,,
\quad
\s^+=
\begin{pmatrix}
	0 & 1 \\ 0 & 0 \\
       \end{pmatrix}\,,
\quad
\s^-=
\begin{pmatrix}
	0 & 0 \\ 1 & 0 \\
       \end{pmatrix}\,.
\end{align*}
\medskip

\noindent{\it Example 2.}\quad Set $\z=\z_2/\z_1$.
We have
\begin{align}
&g^+_{c_2}(\z_2)g^-_{c_1}(\z_1)(I)
=\Delta^{-1}_\z\psi(\z,\al)\bigl(I+q^{\al(\sigma^3_{c_1}+\sigma^3_{c_2})}\bigr)
\label{gep4}\\
&-\frac{1}{2}\bigl(\Delta^{-1}_\z\psi(q\z,\al)+\Delta^{-1}_\z\psi(q^{-1}\z,\al)\bigr)
\left(q^{\al\s^3_{c_1}}+\R_{c_2,c_1}(\z)\bigl(q^{\al\s^3_{c_2}}\bigr)\right)
\nn\\
&-\frac{1}{2}\frac{q^\al+q^{-\al}}{q^\al-q^{-\al}}
\psi(\z,\al)
\left(q^{\al\s^3_{c_1}}-\R_{c_2,c_1}(\z)\bigl(q^{\al\s^3_{c_2}}\bigr)\right)
\nn\\
&+\psi(q\z,\al)\tau^+_{c_1}\tau^-_{c_2}
-\psi(q^{-1}\z,\al)\tau^-_{c_1}\tau^+_{c_2}
-\Bigl(\frac{(q\z)^\al}{q\z-q^{-1}\z^{-1}}
-\frac{(q^{-1}\z)^\al}{q^{-1}\z-q\z^{-1}}\Bigr)\s^-_{c_1}\s^+_{c_2}\,,
\nn\\
&g^-_{c_2}(\z_2)g^+_{c_1}(\z_1)(I)=-\bigl\{
\mathbb{J}\Bigl(g^+_{c_2}(\z_2)g^-_{c_1}(\z_1)(I)\Bigr)
\bigr\}\bigl|_{\al\to -\al}.
\label{gep5}
\end{align}

\bigskip

\subsection{Operators $g^{\epsilon}_{\mathrm{rat},c}(\z)$}\label{sec:annihil}
Our convention and notation used here are 
slightly different from those of \cite{HGSII}, \cite{HGSIII}. 
Let us explain the precise relation between the two.  

For that matter, we need the annihilation operators 
\begin{align*}
x^\pm(\z),\ \bar{x}^+(\z)
~:~ \End V_{[1,n]}\longrightarrow \End V_{[1,n]}\,.
\end{align*}
Define for $X\in \End V_{[1,n]}$ with $s(X)=s$
\begin{align*}
&x^+(\z)(X)
=\int_{\Gamma_0}\psi(\z/\xi,\al)\cdot
\Bigl(\kb(q\xi)+\kb(q^{-1}\xi)\Bigr)(X)
\frac{d\xi^2}{2\pi i \xi^2}\,,
\\
&\bar{x}^+(\z)(X)
=\int_{\Gamma_0}\psi(\z/\xi,\al)\cdot
\kb(\xi)(X)\frac{d\xi^2}{2\pi i \xi^2}\,,
\\
&x^-(\z)(X)=q^{-\al+2s-2}(1-q^{2(\al-2s+1)})\times
\Bigl\{\J \circ x^+(\z)\circ \J(X)\Bigr\}\Bigl|_{\al\to 2s-\al}\,. 
\end{align*}
The contour $\Gamma_0$ encircles $\xi^2=\xi^2_j$ for all $j$. 

Consider the operator 
\begin{align}
&\fb_{\mathrm{rat}}(\z)(X)
=\int_{\Gamma}
\Delta^{-1}\psi(\z/\xi,\al)\cdot
\left(\kb(\xi)-\bar{x}^+(\xi)
-\frac{1}{2}\bigl(x^+(q\xi)+x^+(q^{-1}\xi)\bigr)
\right)(X)
\frac{d\xi^2}{2\pi i \xi^2}\,.
\label{fratbar}
\end{align}
Define further $g^{\pm}_{\mathrm{rat},c}(\z)$
by replacing $\fb(\z)$ by $\fb_{\mathrm{rat}}(\z)$
in the definition \eqref{g+}, \eqref{g-}.
Then the coefficients of $g^{\pm}_{\mathrm{rat},c}(\z)$ 
are rational functions of $\z^2$ (up to an overall power).    

Let us write $\mathbf{b}^*_{II,[1,n]}(\z,\al)$, etc. for the 
operators defined in \cite{HGSII} in the inhomogeneous case.
Then, for $X\in \End V_{[1,n]}$ with $s(X)=s$, we have
\begin{align*}
&g^+_{\mathrm{rat},c}(\z)(X)=\half 
\mathbf{b}^*_{II,[1,n]\cup c}(\z,\al-s-1)(X), 
\\
&g^-_{\mathrm{rat},c}(\z)(X)=\half 
\mathbf{c}^*_{II,[1,n]\cup c}(\z,\al-s+1)(X), 
\\
&g^0_c(\z)(X)=\half \mathbf{t}^*_{II,[1,n]\cup c}(\z,\al-s)(X)\,,
\\
&x^+(\z)(X)=2\mathbf{c}_{II,[1,n]}(\z,\al-s-1)(X),
\\
&x^-(\z)(X)=2\mathbf{b}_{II,[1,n]}(\z,\al-s+1)(X)\,.
\end{align*}
In the right hand side, the interval $[1,n]\cup c$ signifies
$\ds{V_1(\xi_1)\otimes\cdots\otimes
V_n(\xi_n)\otimes V_c(\z)}$. 
The factor $2$ appears because we 
use the normalised trace in this paper.  

\subsection{Dual rqKZ equation}\label{sec:rqKZ1/2}
In addition to $g^0_c(\z),g^\pm_c(\z)$,  
we use the tautological operator
$g^{ \bar{0}}_{c}(\z)$ defined in \eqref{empty}.
Acting with \eqref{g0}-\eqref{g-}
on the unit operator $I$ 
we obtain a set of $4^n$ operators
\begin{align}
g^{\epsilon _n,\cdots,\epsilon_1}(\xi _n,\cdots, \xi _1)
=
g_n^{\epsilon _n}(\xi_n)\cdots g_1^{\epsilon_1}(\xi _1)(I)\,
\quad (\epsilon_1,\cdots,\epsilon_n\in\{0, \bar{0},+,-\}).
\label{g-basis}
\end{align}
For generic $\xi_1,\cdots,\xi_n$ this
gives a basis of $\End V_{[1,n]}$ (see \cite{HGSII}, Lemma 5.1). 

The basis \eqref{g-basis} has a 
distinguished feature that it satisfies the
dual rqKZ equation
\begin{align}
&g^{\epsilon _n,\cdots,\epsilon _j,\cdots,\epsilon_1}
(\xi _n,\cdots,\xi _j,\cdots, \xi _1)
\label{dualrqKZ}
\\&=(-1)^{\epsilon_j}
\bbA_{j,[1,n]}(\xi_1,\cdots,\xi_n)
\circ
g^{\epsilon _n,\cdots,
\overline{\epsilon _j},\cdots,\epsilon_1}
(\xi _n,\cdots,\xi _j q^{-1},\cdots, \xi _1)\,,
\nn
\end{align}
where we set $\overline{\epsilon}=\epsilon$ for $\epsilon=\pm$.  
Operators $\bbA_{j,[1,n]}$ are defined by \eqref{defA}.  

The dual rqKZ equation \eqref{dualrqKZ} is a consequence of 
two identities, which for historical reasons we call 
the first and the second difference equations. 

The first difference equation states that 
\begin{align}
&g^{\epsilon}_{n,[1,n-1]}(\xi_n|\xi_1,\cdots,\xi_{n-1})(X_{[1,n-1]})
\label{1de}\\
&=(-)^\epsilon\bbA_{n,[1,n]}
(\xi_1,\cdots,\xi_n)
\left(
g^{\bar{\epsilon}}_{n,[1,n-1]}(q^{-1}\xi_n|\xi_1,\cdots,\xi_{n-1})(X_{[1,n-1]})
\right)\,
\nn
\end{align}
for $X_{[1,n-1]}$ which acts as identity on the $n$-th component.
For $\epsilon=0, \bar{0}$, this is a restatement of 
\eqref{1dif1}, \eqref{1dift}. The essential case $\epsilon=+$
has been proved in \cite{HGSIII}, Lemma B.2. 

The second difference equation
\begin{align}
&g^{\epsilon}_{n,[1,n-1]}(\xi_n|\xi_1,\cdots,\xi_{n-1})
\circ\bbA_{n-1,[1,n-1]}(\xi_1,\cdots,\xi_{n-1})
\label{2de}\\
&=\bbA_{n-1,[1,n]}(\xi_1,\cdots,\xi_n)
\circ
g^{\epsilon}_{n,[1,n-1]}(\xi_n|\xi_1,\cdots,
q^{-1}\xi_{n-1})
\nn
\end{align}
was not used in \cite{HGSIII}, but can be deduced in a similar (and simpler)
way as for \eqref{1de}.

\subsection{Inductive limit}\label{sec:glimit}

Being constructed through the monodromy matrices, 
$g^\epsilon_c(\z)$ have the $R$ matrix symmetry. 
Namely we have 
\begin{align}
&g^{\epsilon}_{c,[1,n]}
(\z_1|\xi_1,\cdots,\xi_j,\xi_{j+1},\cdots,\xi_n)
\R_{j+1,j}(\xi_{j+1}/\xi_j)
\label{Rsym}\\
&=\R_{j+1,j}(\xi_{j+1}/\xi_j)
g^{\epsilon}_{c,[1,j-1]\cup \{j+1\}\cup\{j\}\cup [j+2,n]}
(\z_1|\xi_1,\cdots,\xi_{j+1},\xi_j,\cdots,\xi_n)\,.
\nn
\end{align}

Operators $g^\epsilon_c(\z)$ enjoy also the following 
right reduction property: 
If $n<N$ and $X_{[1,n]}\in \End V_{[1,n]}$, then 
\begin{align}
g^{\epsilon}_{c,[1,N]}(\z|\xi_1,\cdots,\xi_N)(X_{[1,n]}) 
=\T_{c,[n+1,N]}(\z|\xi_{n+1},\cdots,\xi_N)
g^{\epsilon}_{c,[1,n]}(\z|\xi_1,\cdots,\xi_n)(X_{[1,n]})\,.
\label{g-red}
\end{align}
For $\epsilon=0$ this is immediate.  
The non-trivial case $\epsilon=+$ follows from Lemma 3.7 in \cite{HGSII}. 
In \cite{HGSII} we considered $g^\epsilon_{\mathrm{rat},c}(\z)$, but this equation is equally valid for 
$g^\epsilon_{c}(\z)$ since they
differ only by annihilation operators.

From \eqref{g-red} we can deduce 
the reduction property for the composition \eqref{g-comp}:
\begin{align}
& g^{\epsilon_2}_{c_2,[1,N]\cup c_1}(\z_2|\xi_1,\cdots,\xi_N,\z_1)
g^{\epsilon_1}_{c_1,[1,N]}(\z_1|\xi_1,\cdots,\xi_N)(X_{[1,n]}) 
\label{gg-comp}
\\
&=g^{\epsilon_2}_{c_2,[1,N]\cup c_1}(\z_2|\xi_1,\cdots,\xi_N,\z_1)
\T_{c_1,[n+1,N]}(\z_1|\xi_{n+1},\cdots,\xi_N)
g^{\epsilon_1}_{c_1,[1,n]}(\z_1|\xi_1,\cdots,\xi_n)(X_{[1,n]}) 
\nn\\
&=\T_{c_1,[n+1,N]}(\z_1|\xi_{n+1},\cdots,\xi_N)
g^{\epsilon_2}_{c_2,[1,n]\cup c_1\cup[n+1,N]}
(\z_2|\xi_1,\cdots,\xi_n,\z_1,\xi_{n+1},\cdots,\xi_N)
\nn\\
&\quad \times 
g^{\epsilon_1}_{c_1,[1,n]}(\z_1|\xi_1,\cdots,\xi_n)(X_{[1,n]}) 
\nn\\
&=\T_{c_1,[n+1,N]}(\z_1|\xi_{n+1},\cdots,\xi_N)
\T_{c_2,[n+1,N]}(\z_2|\xi_{n+1},\cdots,\xi_N)
\nn\\
&\quad\times 
g^{\epsilon_2}_{c_2,[1,n]\cup c_1}(\z_2|\xi_1,\cdots,\xi_n,\z_1)
g^{\epsilon_1}_{c_1,[1,n]}(\z_1|\xi_1,\cdots,\xi_n)(X_{[1,n]}) \,.
\nn
\end{align}
In the third line we used the $R$ matrix symmetry \eqref{Rsym}.

Now let us fix $\{\xi_j\}_{j=1}^\infty$ in such a way that 
$\xi_j=1$ for $j\gg 1$. 
The right hand side of \eqref{g-red} can be rewritten as 
\begin{align*}
\check{\R}_{c,N}(\z/\xi_N) \cdots \check{\R}_{n+2,n+1}(\z/\xi_{n+1})
g^\epsilon_{n+1,[1,n]}(\z|\xi_1,\cdots,\xi_n)(X_{[1,n]})\,.
\end{align*}
Therefore, by the same argument as in Section \ref{sec:gen}, we see that 
the inductive limit
\begin{align}
g^\epsilon(\z)(X_{[1,n]})
:=\lim_{N\to\infty} g^{\epsilon}_{c,[1,N]}(\z|\xi_1,\cdots,\xi_n)(X_{[1,n]}) 
\label{g-indlim}
\end{align}
is well defined as a formal series in $\z^2-1$. 
For any given $p$ 
\begin{align*}
&g^\epsilon(\z)(X_{[1,n]}) \\
&\quad\equiv
\Tr_c\left\{
\T_{c,[n+1,N]}(\z|\xi_{n+1},\cdots,\xi_N)
g^{\epsilon}_{c,[1,n]}(\z|\xi_1,\cdots,\xi_n)(X_{[1,n]})
\right\}
  \mod (\z^2-1)^p\,,
\end{align*}
holds for $N$ large enough. 

Letting $N\to\infty$ in \eqref{gg-comp}, 
we conclude that
\begin{align}
& g^{\epsilon_2}(\z_2)g^{\epsilon_1}(\z_1)(X_{[1,n]}) 
\label{g-ind-comp}\\
&\quad=
\Tr_{c_1,c_2}\left\{\T_{c_1,[n+1,\infty)}(\z_1) \T_{c_2,[n+1,\infty)}(\z_2)
g^{\epsilon_2}_{c_2}(\z_2)g^{\epsilon_1}_{c_1}(\z_1)(X_{[1,n]})
\right\} \, .
\nn
\end{align}
Obviously we have in general 
\begin{align}
& g^{\epsilon_k}(\z_k)\cdots g^{\epsilon_1}(\z_1)(X_{[1,n]}) \label{Rdoll}
\\
&\quad=
\Tr_{c_1,\cdots,c_k}
\left\{\T_{c_1,[n+1,\infty)}(\z_1)\cdots \T_{c_k,[n+1,\infty)}(\z_k)
g^{\epsilon_k}_{c_k}(\z_k)\cdots g^{\epsilon_1}_{c_1}(\z_1)(X_{[1,n]})
\right\} \, .\nn
\end{align}

Let us give a remark about the commutation relations.  
Rewriting the right hand side of \eqref{g-ind-comp} as
\begin{align*}
\Tr_{c_1,c_2}
\left\{\T_{c_2,[n+1,\infty)}(\z_2)\T_{c_1,[n+1,\infty)}(\z_1) 
\R_{c_1,c_2}(\z_1/\z_2)
g^{\epsilon_2}_{c_2}(\z_2)g^{\epsilon_1}_{c_1}(\z_1)(X_{[1,n]})
\right\} \,, 
\end{align*}
and comparing it with 
$g^{\epsilon_1}(\z_1)g^{\epsilon_2}(\z_2)(X_{[1,n]})$, 
we see that 
\begin{align}
g^{\epsilon_1}(\z_1)g^{\epsilon_2}(\z_2) 
=(-)^{\epsilon_1\epsilon_2}
g^{\epsilon_2}(\z_2) g^{\epsilon_1}(\z_1) 
\label{g-ind-comm}
\end{align}
is equivalent to 
\begin{align}
g_{c_1}^{\epsilon_1}(\z_1)g_{c_2}^{\epsilon_2}(\z_2)
=(-)^{\epsilon_1\epsilon_2}
\R_{c_1,c_2}(\z_1/\z_2)
g_{c_2}^{\epsilon_2}(\z_2)g_{c_1}^{\epsilon_1}(\z_1)\,. 
\label{g-comm}
\end{align}

In the main text, we extend 
\eqref{g-indlim} further to define operators acting on the 
space $\mathcal{W}^{(\al)}$ of quasi-local operators,
\begin{align*}
x^{\epsilon*}(\z)\bigl(q^{(\al-s+\epsilon)H(0)}\mathcal{O}\bigr)
= q^{(\al-s)H(0)}
g^\epsilon(\z)\bigl(\mathcal{O}_{[1,n]}\bigr)\,,
\end{align*}
where $\mathcal{O}$ is supported on $[1,n]$. 
The relation \eqref{g-ind-comm} is nothing but  the (anti-)commutativity 
between $b^*(\z)=x^{+*}(\z)$, $c^*(\z)=x^{-*}(\z)$ and $t^*(\z)=x^{0*}(\z)$. 
In this latter form the commutation relations have been proved in \cite{COMM} 
by an indirect argument based on the results of \cite{HGSIII}.

\subsection{Expectation values}\label{sec:VEV1/2}
The main result of \cite{HGSIII} states that the expectation values
of quasi-local operators in the fermionic basis are expressed as determinants
of the simplest one:
\begin{align*}
\omega(\z,\xi)=
Z^\kappa_{\mathbf{n}} 
\Bigl\{b^*(\z)c^*(\xi)\bigl(q^{\al H(0)}\bigr)\Bigr\}\,.
\end{align*}
The precise relation between this function and the one in \cite{HGSIII} is as follows.

We assume that in the Matsubara direction 
there are $\mathbf{n}$ number of rows 
corresponding to representations of dimension $d_{\mathbf{m}}$ 
and evaluation parameters $\taub_{\mathbf{m}}$, 
and set  
\begin{align*}
a(\z)=\prod_{\mathbf{m}=1}^\mathbf{n}\bigl(
q^{d_\mathbf{m}}\z^2/\taub^2_{\mathbf{m}}-1\bigr),
\quad  
d(\z)=\prod_{\mathbf{m}=1}^\mathbf{n}\bigl(q^{2-d_\mathbf{m}}
\z^2/\taub^2_{\mathbf{m}}-1\bigr)\,.
\end{align*}
Denote by $\omega_{III}(\z,\xi)$ 
the function defined in \cite{HGSIII}, eq.(7.2). 
Then we have 
\begin{align}
\omega(\z,\xi)=\frac{1}{4}\omega_{III}(\z,\xi)
+\frac{1}{4}\overline{D}_{\z}\overline{D}_{\xi}\Delta^{-1}_{\z}\psi(\z/\xi,\al)
\end{align}
where 
\begin{align*}
&\overline{D}_\z f(\z)=f(\z q)+f(\z q^{-1})-2\rho(\z)f(\z)\,,
\\
&\rho(\z)=\frac{T(\z,\kappa+\al)}{T(\z,\kappa)}\,.
\end{align*}
The function  $\omega(\z,\xi)$ has singularities at 
$\z=\xi q^{\pm1}$, with the singular part given by 
\begin{align*}
&\frac{1}{4}\omega_{III,\mathrm{symm}}(\z,\xi)
+\frac{1}{4}\overline{D}_{\z}\overline{D}_{\xi}\Delta^{-1}_{\z}\psi(\z/\xi,\al)
\\
&=\bigl(1+\rho(\z)\rho(\xi)\bigr)\Delta^{-1}_{\z}\psi(\z/\xi,\al)
-\rho(\xi)\Delta^{-1}_{\z}\psi(q\z/\xi,\al)-\rho(\z)\Delta^{-1}_{\z}\psi(q^{-1}\z/\xi,\al)
\\
&+\frac{1}{4}
\frac{a(\xi)d(\z)}{T(\xi,\kappa)T(\z,\kappa)}\psi(q\z/\xi,\al)
-\frac{1}{4}
\frac{a(\z)d(\xi)}{T(\z,\kappa)T(\xi,\kappa)}\psi(q^{-1}\z/\xi,\al)\,.
\end{align*}

\setcounter{section}{1}

\section{Admissibility of $g^{(2,2)}_{c_1,c_2}(\z)$}\label{sec:appB}
We show here the admissibility of $g^{(2,2)}_{c_1,c_2}(\z)$. 
Consider the combination 
\begin{align*}
G(\z_2,\z_1)=g^{+}_{c_2}(\z_2) g^{-}_{c_1}(\z_1)+g^{-}_{c_2}(\z_2) g^{+}_{c_1}(\z_1)\,.
\end{align*}
The possible poles of $G(\z_2,\z_1)$ in $\z_2/\z_1$ are at the shifted diagonal $\z_2/\z_1=q^m$. 
Among them, $m=0$ is known to be absent \cite{HGSII}.  
In addition, the commutation relation implies
\begin{align}
G(\z,\z)&=0\,.\label{zero}
\end{align}
On the other hand, the dual rqKZ equation reads
\begin{align}
G(\z_2,\z_1)&
=-R_{c_2,c_1}(\z_2/\z_1)
q^{\al H_{c_2}}\theta_{c_2}
\Bigl(G(\z_2q^{-1},\z_1)
\theta_{c_2}
\bigl(R_{c_2,c_1}(\z_2/\z_1)^{-1}\bigr)
\Bigr)\,,
\label{drqkz1}\\
&=-q^{\al H_{c_1}}\theta_{c_1}\Bigl(
\theta_{c_1}\bigl(R_{c_1,c_2}(\z_1/\z_2)^{-1}\bigr)
G(\z_2,\z_1q^{-1})\Bigr)R_{c_1,c_2}(\z_1/\z_2)\,.\nn
\end{align}
Combining \eqref{zero}, \eqref{drqkz1} we find that $G(\z_2,\z_1)$ is in fact regular at $\z_2/\z_1=q^m$
for all $m\in \Z$. Moreover the first equation of \eqref{drqkz1} implies that 
$G(q\z,\z)=\mathcal{P}^+_{c_1,c_2}G(q\z,\z)$, showing that $G(q\z,\z)$ is admissible. 
\bigskip


\bigskip

{\it Acknowledgements.}\quad
\medskip

MJ and TM would like to thank Andreas Kl{\" u}mper and 
Junji Suzuki for discussions. 

Research of MJ is supported by the Grant-in-Aid for Scientific 
Research B-23340039.
Research of TM is supported by the Grant-in-Aid for Scientific 
Research B-22340031.
Research of FS is supported  by DIADEMS program (ANR) 
contract number BLAN012004.


\end{document}